\documentclass[twocolumn,superscriptaddress,amsmath,amssymb]{revtex4}
\usepackage{amsmath}
\usepackage{amssymb}
\usepackage{babel}
\usepackage{graphicx}
\usepackage{dcolumn}
\usepackage{bm}
\usepackage{hyperref}
\usepackage{mathrsfs} 
\usepackage{gensymb} 
\usepackage[absolute,overlay]{textpos}
\usepackage{upgreek} 
\usepackage{textcomp} 
\usepackage{xcolor}
\usepackage[caption=false]{subfig}

\begin{document}

\title{Nematic fluctuations mediated superconductivity revealed by anisotropic strain in Ba(Fe$_{1-x}$Co$_x$)$_2$As$_2$}

\author{Jean-Côme Philippe}
\author{Alexis Lespinas}
\author{Jimmy Faria}
\affiliation{Université Paris Cité, Matériaux et Phénomènes Quantiques, UMR CNRS 7162, Bâtiment Condorcet, 75205 Paris Cedex 13, France}
\author{Anne Forget}
\author{Dorothée Colson}
\affiliation{Service de Physique de l'Etat Condensé, DSM/DRECAM/SPEC, CEA Saclay, Gif-sur-Yvette, 91191, France}
\author{Sarah Houver}
\author{Maximilien Cazayous}
\author{Alain Sacuto}
\affiliation{Université Paris Cité, Matériaux et Phénomènes Quantiques, UMR CNRS 7162, Bâtiment Condorcet, 75205 Paris Cedex 13, France}
\author{Indranil Paul}
\affiliation{Université Paris Cité, Matériaux et Phénomènes Quantiques, UMR CNRS 7162, Bâtiment Condorcet, 75205 Paris Cedex 13, France}
\author{Yann Gallais}
\email{yann.gallais@u-paris.fr}
\affiliation{Université Paris Cité, Matériaux et Phénomènes Quantiques, UMR CNRS 7162, Bâtiment Condorcet, 75205 Paris Cedex 13, France}

\date{\today}

\begin{abstract}
Anisotropic strain is an external field capable of selectively addressing the role of nematic fluctuations in promoting superconductivity. We demonstrate this using polarization-resolved elasto-Raman scattering by probing the evolution of nematic fluctuations under strain in the normal and superconducting state of the paradigmatic iron-based superconductor Ba(Fe$_{1-x}$Co$_x$)$_2$As$_2$. In the parent compound BaFe$_2$As$_2$ we observe a strain-induced suppression of the nematic susceptibility which follows the expected behavior of an Ising order parameter under a symmetry breaking field. For the superconducting compound, the suppression of the nematic susceptibility correlates with the decrease of the critical temperature $T_c$, indicating a significant contribution of nematic fluctuations to electron pairing. Our results validates theoretical scenarios of enhanced $T_c$ near a nematic quantum critical point.
\end{abstract}

\maketitle

In many iron-based superconductors (FeSC), such as Ba(Fe$_{1-x}$Co$_x$)$_2$As$_2$ (denoted thereafter as Co:Ba122), superconductivity (SC) occurs around the end point of stripe-like antiferromagnetic (AF) and nematic phases, suggesting a link between SC and critical fluctuations associated with the proximity of a nematic or magnetic quantum critical point (QCP) ~\cite{shibauchi_quantum_2014, fernandes_what_2014,kuo_ubiquitous_2016}. Initially SC was believed to result from magnetic fluctuations~\cite{kuroki_unconventional_2008,mazin_unconventional_2008, scalapino_common_2012}, but following the observation of strong nematic fluctuations through various probes~\cite{chu_divergent_2012,yoshizawa_structural_2012,gallais_observation_2013, gallais_charge_2016,bohmer_nematic_2014, bohmer_electronic_2016,ikeda_elastocaloric_2020} nematic degrees of freedom, which break the lattice rotation symmetry while preserving its translation symmetry, have been envisioned as a possible alternative source for the enhancement of the superconducting critical temperature $T_c$~\cite{yamase_superconductivity_2013,metlitski_cooper_2015,lederer_enhancement_2015, maier_pairing_2014,labat_pairing_2017,eckberg_sixfold_2019,lederer_tests_2020}. Unfortunately, magnetic and nematic fluctuations are difficult to disentangle in most FeSC since both phases lie in close proximity. To assess the role of critical nematic fluctuations in enhancing $T_c$, it is essential to correlate $T_c$ with nematic fluctuations close to the nematic QCP using a stimulus that selectively tune nematic degrees of freedom.
\par
In this context, anisotropic strain provides an appealing tuning parameter to disentangle the role of magnetic and nematic degrees of freedom in promoting SC because it directly couples to the nematic order parameter $\phi_{nem}$ provided it has the relevant symmetry, the $B_{2g}$ representation in the case of FeSC ~\cite{dhital_effect_2012,dhital_evolution_2014, mirri_electrodynamic_2016, tam_uniaxial_2017, kissikov_uniaxial_2018, pfau_anisotropic_2021}. This was demonstrated in the weak-field limit via elastoresistivity measurements which allowed the extraction of the nematic susceptibility $\chi_{nem}$~\cite{chu_divergent_2012}. In the strong field limit, anisotropic strain can also be used as a selective tool to induce or enhance nematic order while leaving the magnetic order comparatively less affected \cite{kissikov_uniaxial_2018,sanchez_transportstructural_2021,philippe_elasto-raman_2022}. This is because an uniform ($q$=0) anisotropic strain couples linearly to $\phi_{nem}$, but to the finite wavevector $Q$ magnetic order parameter only indirectly via higher orders couplings. Recently, Malinowski et al.~\cite{malinowski_suppression_2020} have revealed in Co:Ba122 a large suppression of $T_c$ under anisotropic strain near the QCP, suggesting an intimate link between SC and nematicity. However, transport measurements cannot probe the nematic fluctuations in the superconducting state, so that the precise link between nematic fluctuations and SC remains to be established in this material. 

Here we report an elasto-Raman spectroscopy study on Co:Ba122 establishing a link between nematic fluctuations and $T_c$ under anisotropic strain. In the parent compound Ba122 the effect of strain on nematic fluctuations displays the hallmarks of the susceptibility of an Ising order parameter under a symmetry breaking field. A strong and symmetric reduction of $\chi_{nem}$ with strain is observed near the structural transition temperature $T_s$ resulting in a significant suppression of its temperature dependence. For the superconducting compound, a similar reduction of $\chi_{nem}$ is observed under strain in both the superconducting and normal states. We further show that the reduction of $\chi_{nem}$ scales linearly with $T_c$, indicating a link between $T_c$ and nematic fluctuations at optimal doping. Our results showcase a dominant role for nematic fluctuations in boosting $T_c$ in Co:Ba122.

Two Co:Ba122 single crystals were investigated. Samples from the same batch were previously studied by transport measurements, from which superconducting $T_c$, nematic $T_s$ and AF $T_N$ transition temperatures were determined, and by Raman scattering under nominally zero strain \cite{rullier-albenque_hall_2009,chauviere_doping_2009,chauviere_raman_2011,gallais_observation_2013,gallais_nematic_2016}. The first crystal is the parent compound, BaFe$_2$As$_2$ ($x$=0), which displays a simultaneous magnetic (from paramagnetic to AF) and structural (from tetragonal to orthorhombic) transition at $T_{s/N} = 138$~K and no superconducting state. The second crystal is close to the optimal doping and to the nematic QCP, with $x = 0.07$, $T_c = 24$~K as determined by SQUID magnetometry on the same crystal. It presents no magnetic order and remains tetragonal down to low temperatures. We use an uniaxial piezoelectric cell (CS130 from Razorbill Instruments) to apply both compressive and tensile stress upon a sample glued between two mounting plates. The stress is applied along the long dimension being the [110] direction of the usual two Fe unit cell (that is along the Fe-Fe bonds and denoted $x'$ hereafter), resulting in an anisotropic $B_{2g}$ strain which couples to the $B_{2g}$ nematic order parameter. All the Raman spectra have been corrected for the Bose factor and the instrumental spectral response. They are thus proportional to the imaginary part of the Raman response function $\chi(\omega,T)$. Additional experimental details are given in the Supplemental Material~\cite{SI} and in Ref.~\cite{philippe_elasto-raman_2022}, where elasto-Raman phonon spectra obtained on the $x=0$ sample were already presented.

In the following, we consider an ($x'y'z'$) frame with the $x'$ axis parallel to the applied stress direction. We define $\epsilon_{x^{\prime} x^{\prime}}^{nom}$ as the nominal applied strain along the $x'$ direction monitored in situ through the measured displacement of the mounting plates. The strain experienced by the sample along the stress direction $\epsilon_{x^{\prime} x^{\prime}}$ differs from $\epsilon_{x^{\prime} x^{\prime}}^{nom}$ because of the imperfect strain transmission through the epoxy glue \cite{philippe_elasto-raman_2022}. Also due to finite Poisson ratio, the actual strain is triaxial, and $\epsilon_{x^{\prime} x^{\prime}}$ can be decomposed under both two isotropic $A_{1g}$ and one anisotropic $B_{2g}$ components~\cite{ikeda_symmetric_2018}. Throughout the paper, the data will be shown as a function of $\epsilon_{x^{\prime} x^{\prime}}^{nom}$, but the effects of strain transmission will be taken into account when comparing the data on different samples. 

\begin{figure}[h!]
    \centering
    \includegraphics[width=0.49\textwidth]{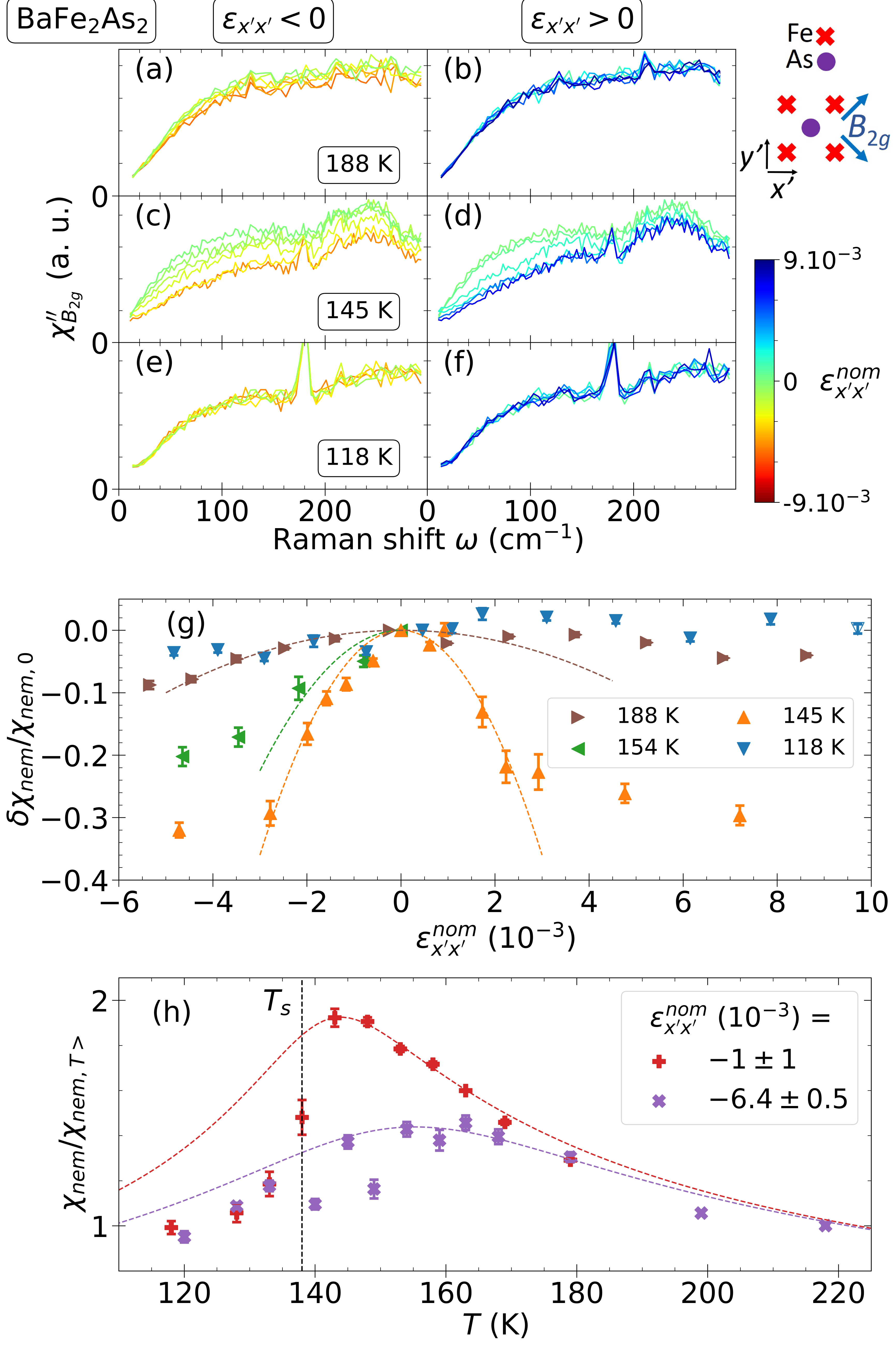}
    \caption{Strain effect on the nematic ($B_{2g}$) Raman response for the parent compound BaFe$_2$As$_2$ ($x=0$). (a) to (f) Raman response obtained at 188 ((a)-(b)), 145 ((c)-(d)) and 118~K ((e)-(f)), upon negative ((a)-(c)-(e)) or positive ((b)-(d)-(f)) strain. The sketch in the upper right displays the directions of the incident and scattered polarizations (blue arrows) with respect to the Fe atom square lattice. (g) Relative variation of $\chi_{nem}$ upon strain with respect to the zero strain susceptibility at fixed temperature. No data was obtained at 154~K under tensile strain as the sample broke during the experiment. The dashed lines are guides to the eye using quadratic law (from Eq.~\eqref{eqDeltaChiEpsQuadratique}). (h) Temperature dependence of $\chi_{nem}$ at fixed strain, renormalized by the high-temperature susceptibility denoted $\chi_{nem,T>}$ (taken here at 220~K under large strain). The dashed lines correspond to the expected theoretical behavior of $\chi_{nem}(T)$ under weak and strong external fields~\cite{SI}.
    For (g) and (h), the error bars take into account the uncertainty on the Drude-like low energy extrapolation of the spectra but not the uncertainty related to changes in spot or optical alignment that likely dominate the uncertainty in the temperature dependence ~\cite{SI}}.
    \label{fig1}
\end{figure}

All Raman spectra reported here were obtained using crossed polarizations along the principal directions of the 2-Fe unit cell ($xy$ geometry), and thus probe the $B_{2g}$ representation of the $D_{4h}$ point group (see Fig. \ref{fig1}). In this geometry the measured Raman response, denoted $\chi_{B_{2g}}^{\prime\prime}$ in the following, probes the dynamical nematic fluctuations associated to the order parameter $\phi_{nem}$.
They can be related to the nematic static susceptibility $\chi_{nem}$ through~\cite{gallais_charge_2016}:  
\begin{equation}
\chi_{nem} = \int_{0}^{\infty} \frac{\chi_{B_{2g}}^{\prime\prime}(\omega)}{\omega}d\omega
\label{eq:chiNem}
\end{equation}

We first present the results concerning the parent compound (Figure~\ref{fig1}). Just above $T_s$ at 145~K, the $B_{2g}$ Raman response is strongly reduced below 300~cm$^{-1}$ both under positive and negative strains. The suppression occurs also at larger temperature (188~K), but is less pronounced. Below $T_s$ at 118~K, $\chi_{B_{2g}}^{\prime\prime}$ changes very little with strain. At lower temperature Raman fingerprints of AF spin-density-wave (SDW) induced gaps indicate a strengthening of the magnetic order under strain (shown in Fig. S6 of the Supplemental Material \cite{SI}), in agreement with previous Nuclear Magnetic Resonance measurements \cite{chauviere_raman_2011,kissikov_uniaxial_2018}.

With Equation~\eqref{eq:chiNem}, the decrease of $\chi_{B_{2g}}^{\prime\prime}$ with strain can be interpreted as a suppression of the nematic fluctuations. Before computing $\chi_{nem}$ the experimental spectra were extrapolated down to $\omega = 0$ using a Drude lineshape~ (see Fig. S5 of the Supplemental Material \cite{SI}). 
Because the Raman response does not decrease at high energy~\cite{gallais_observation_2013} we use an upper cut-off frequency $\omega_c = 400$~cm$^{-1}$ in the computation of $\chi_{nem}$, chosen as the upper limit above which no strain effect is observed at all temperatures. 
The impact of strain on $\chi_{nem}$ for four temperatures around $T_s$ is displayed in Fig.~\ref{fig1}-(g): we define $\chi_{nem,0}$ as the nematic susceptibility at zero strain, and $\delta\chi_{nem}(\epsilon) = \chi_{nem}(\epsilon) - \chi_{nem,0}$. Above $T_s$, at T=145K, we obtain a clear and symmetric in $\epsilon_{x'x'}^{nom}$ suppression of $\chi_{nem}$ with strain. The suppression significantly weakens as $T$ increases. Below $T_s$, at 118~K, $\chi_{nem}$ hardly displays any strain dependence. At 145~K a clear saturation behavior is observed under high tensile and compressive strains. The temperature dependence of $\chi_{nem}$ under constant applied strain is depicted in Fig.~\ref{fig1}-(h). At high strain, the Curie-Weiss-like divergence of $\chi_{nem,0}(T)$ at $T_s$ observed at low strain is strongly suppressed, and the maximum of $\chi_{nem}(T)$ is shifted to a higher temperature. 
\par
The behavior of $\chi_{nem}$ under strain can be rationalized in the simple picture of an Ising nematic order parameter coupled to a symmetry breaking field. Using a Landau expansion of the free energy in a mean-field framework in both the nematic and elastic order parameters $\phi_{nem}$ and $\epsilon_{xy}$, we obtain the following variation of $\delta\chi_{nem}$ with strain ~\cite{SI}:
\begin{equation}
    \frac{\delta\chi_{nem}}{\chi_{nem,0}} = 3b[\phi_{nem}^2(\epsilon_{x^{\prime} x^{\prime}}=0) - \phi_{nem}^2(\epsilon_{x^{\prime} x^{\prime}})]\chi_{nem}(\epsilon_{x^{\prime} x^{\prime}})
\end{equation}
with $b > 0$ the prefactor of the quartic $\phi_{nem}$ term in the free energy expression, which we consider strain-independent. Restricting ourselves to the low strain regime and to $T > T_s$, to quadratic order in the applied strain we obtain~\cite{SI}:
\begin{equation}
\frac{\delta\chi_{nem}}{\chi_{nem,0}}  \approx - 12 \left( \frac{C_{A}^{}}{C_A + \tilde{C}_{66}} \right)^2 b \lambda^2 \chi_{nem,0}^{3} \epsilon_{x^{\prime} x^{\prime}}^2
\label{eqDeltaChiEpsQuadratique}
\end{equation}
with $\lambda > 0$ the nemato-elastic coupling constant. $C_A=C_{11}+C_{12}$ and $\tilde{C}_{66}$ are the in-plane isotropic and shear elastic modulus which we define using Voigt notation \cite{SI}.

Equation~\eqref{eqDeltaChiEpsQuadratique} qualitatively reproduces two key experimental findings of Figure~\ref{fig1}. First, $\chi_{nem}$ should indeed decrease upon strain, following a symmetric quadratic behavior with $\epsilon_{x^{\prime} x^{\prime}}$ at low strain. Second, since $\chi_{nem,0}$ increases as $T$ approaches $T_s$ \cite{gallais_observation_2013,kretzschmar_critical_2016}, the suppression of $\chi_{nem}$ with $\epsilon_{x^{\prime} x^{\prime}}$ should be larger close to $T_s$ in agreement with our results. As shown in Figure~\ref{fig1}(h) this picture also captures the temperature dependence of $\chi_{nem}$ under strain, and in particular the upward shift of its maximum at strong strain~\cite{SI}. Significant deviations below $T_s$ at low strain are likely due to higher order terms in the free energy and/or the onset of AF order which are not included in the Landau expansion. Overall the behavior of $\chi_{nem}$ for the parent compound thus fits very well the expectations of an Ising-nematic order parameter under a symmetry breaking field.
\par
\begin{figure}[h!]
    \centering
    \includegraphics[width=0.49\textwidth]{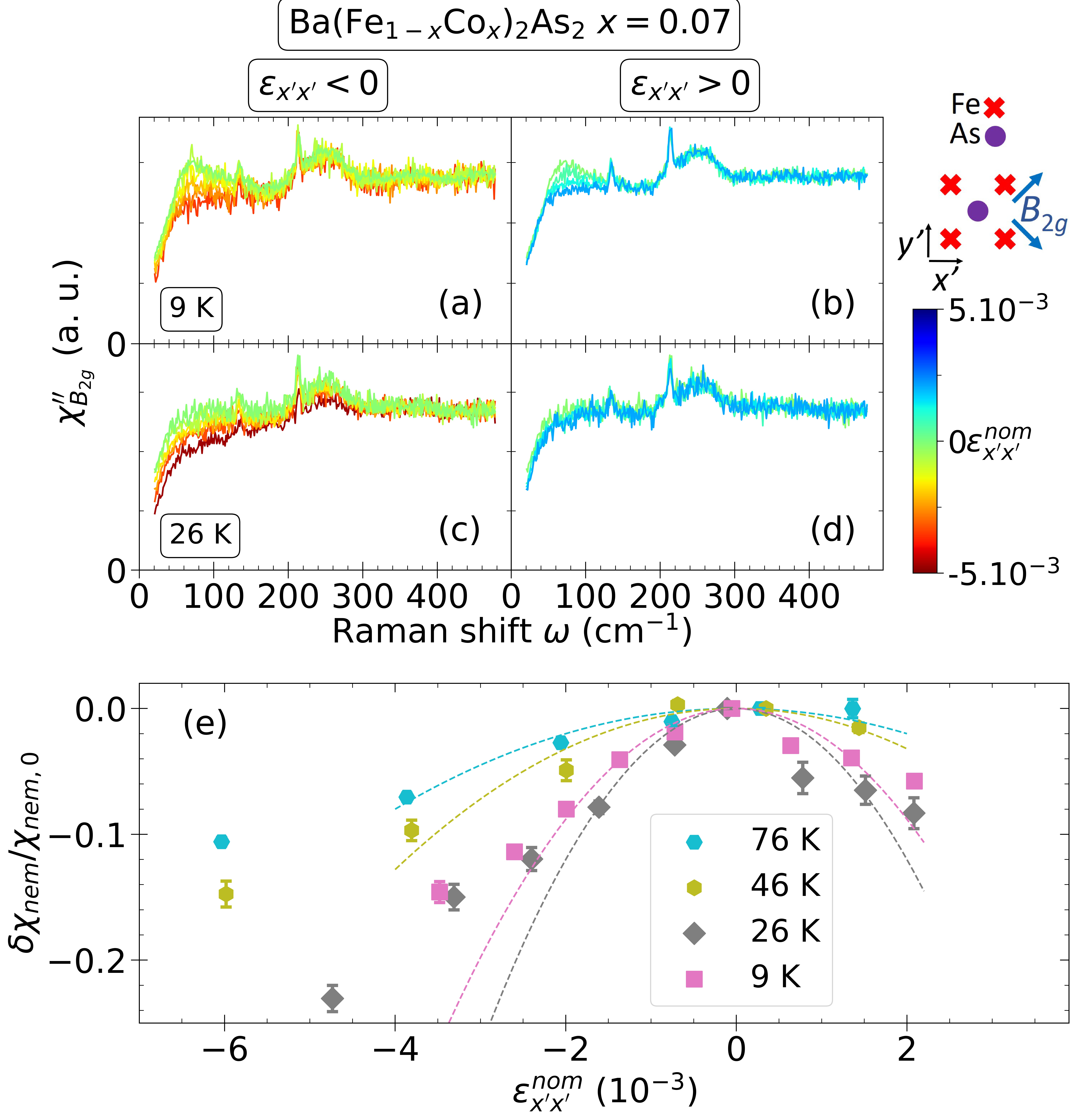}
    \caption{Strain effect on the nematic ($B_{2g}$) Raman response for Ba(Fe$_{1-x}$Co$_x$As)$_2$ (x=0.07). (a)-(d) Raman response obtained at 9 ((a)-(b)) and 26~K ((c)-(d)), upon negative ((a)-(c)) or positive ((b)-(d)) strain. (e) Relative variation of $\chi_{nem}$ upon strain with respect to the zero strain susceptibility at fixed temperature. The dashed lines are guides to the eye obtained by following quadratic laws. The error bars take into account the uncertainty on the low energy extrapolation of the spectra~\cite{SI}. For the 9~K spectra, the maximum of $\chi_{nem}$ occurs at a nominally small compressive strain of approximately -0.5$\times$10$^{-3}$, whereas for all other probed temperatures for the two samples the maximum is located very close to nominally zero strain. The offset at 9~K is likely due to plastic deformation in the epoxy glue which occurred just before this measurement~\cite{barber_thesis_2018}. Therefore the 9~K data have been shifted by an offset of +0.5$\times$10$^{-3}$ on $\epsilon_{x^{\prime}x^{\prime}}^{nom}$.}
    \label{fig2}
\end{figure}

We now move to the results on the doped compound ($x=0.07$) (Fig.~\ref{fig2}) whose composition lies slightly beyond the nematic QCP, located at $x\sim 0.065$ \cite{rullier-albenque_hall_2009,gallais_charge_2016}. At 9~K, below $T_c$, the unstrained Raman response shows a relatively broad superconductivity-induced peak centered around 75~cm$^{-1}$. This was observed before~\cite{muschler_band-_nodate,chauviere_impact_2010}, and interpreted as a nematic resonance mode where the usual Raman pair-breaking peak at twice the SC gap energy, $2\Delta$, is replaced by a collective mode below $2\Delta$ because of significant nematic correlations in the SC state~\cite{gallais_nematic_2016,thorsmolle_critical_2016,adachi_superconducting_2020}. Under compressive and tensile strains, the Raman response in the $B_{2g}$ channel is strongly reduced at 9~K and 26~K, indicating a suppression of the nematic fluctuations below and just above $T_c$. By contrast the spectra in the complementary $B_{1g}$ symmetry channel (displayed in Fig. S4 of the Supplemental Material \cite{SI}) hardly show any strain dependence. The strain variation of $\chi_{nem}$ (Fig.~\ref{fig2} - (e)) at different temperatures shows a qualitatively similar behavior as for $x=0$: a symmetric suppression of $\chi_{nem}$ with respect to strain, and a weakening of the strain dependence at higher temperature which follows the behavior of $\chi_{nem}$ under zero applied strain \cite{gallais_observation_2013} as expected from Equation \ref{eqDeltaChiEpsQuadratique}. 

We stress that the suppression of $\chi_{nem}$ in the superconducting state cannot be simply explained by a reduction of the SC gap magnitude by strain. Indeed in the absence of any nematic correlations, $\chi_{nem}$ is just the normal state density of state at the Fermi energy weighted by the $B_{2g}$ Raman vertex. It is therefore independent of the SC gap energy \cite{labat_variation_2020}, and the suppression of $\chi_{nem}$ must be linked to the suppression of nematic fluctuations in the SC state \cite{SI}. Moreover, the strain-induced suppression of the SC peak intensity further strengthens the nematic resonance hypothesis: the emergence of the SC peak below $T_c$ is closely linked to the presence of significant nematic correlations in the SC state close to the nematic QCP, as evidenced by the approximate scaling between the SC peak weight and $\chi_{nem}$ under strain (see Fig. S2 of the Supplemental Material \cite{SI}). 

\begin{figure}
    \centering
    \includegraphics[width=0.49\textwidth]{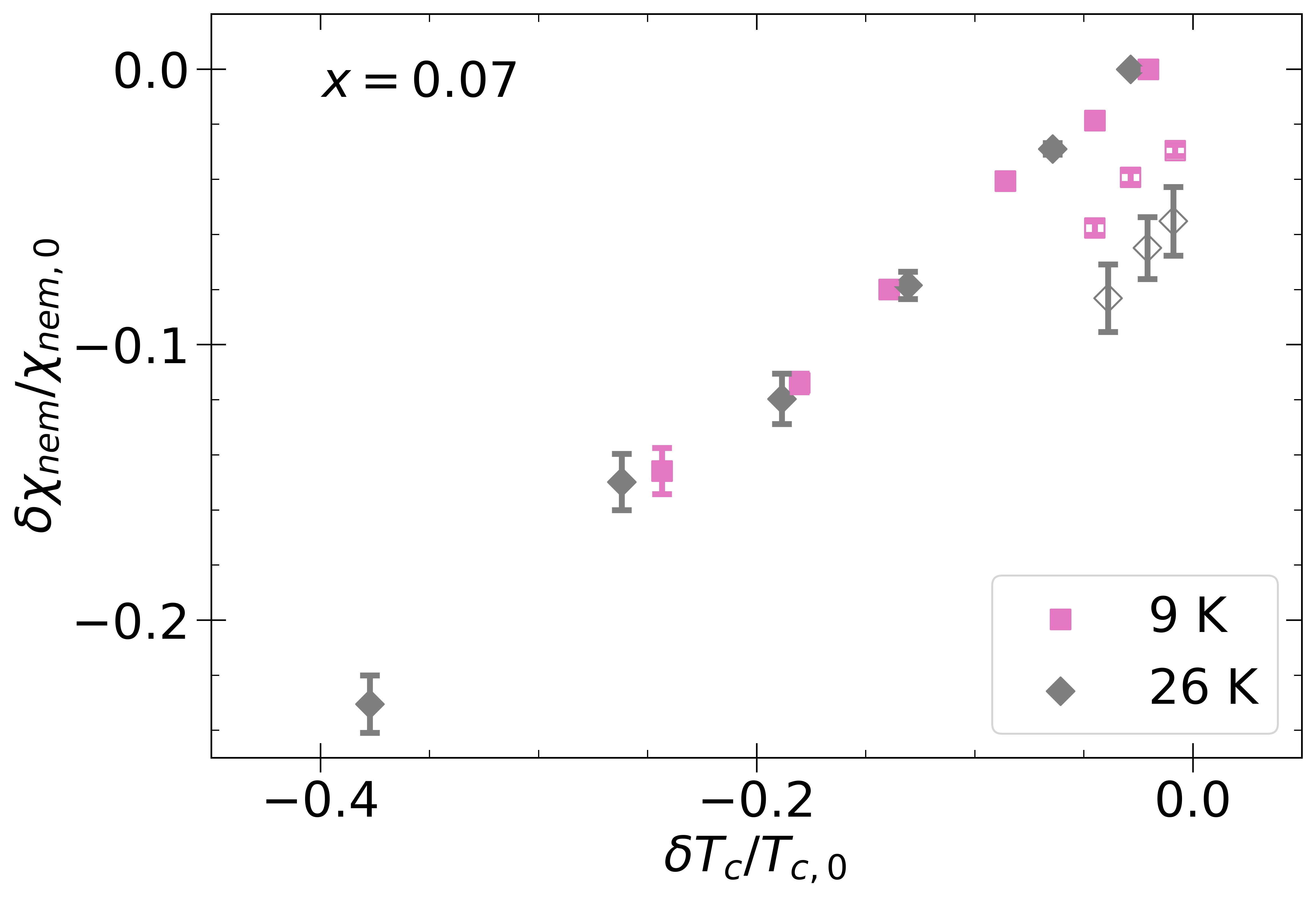}
    \caption{Scaling between the suppression of $\chi_{nem}$ in the superconducting state measured through elasto-Raman spectroscopy and the fall of $T_{c}$ measured in transport on a crystal with similar doping $x$. The $T_c(\epsilon_{x^{\prime} x^{\prime}}^{nom})$ data are taken from Ref.~\cite{malinowski_suppression_2020}. The estimated strain transmission ratio have been taken into account for both data sets \cite{SI,malinowski_suppression_2020}. The empty symbols stand for the data under positive strain, the filled ones under negative strain. The offset between positive and negative strain data points is partly attributed to uncertainties in the exact location of the zero strain state, but could also possibly reflect an intrinsic asymmetry between tensile and compressive strain~\cite{SI}.}
    \label{fig3}
\end{figure}

\par
Our results suggest a connection between the suppression of $\chi_{nem}$ in the superconducting state and the rapid fall of $T_{c}$ observed in transport measurements on a similarly doped sample~\cite{malinowski_suppression_2020}.  In the case where SC pairing is promoted by nematic fluctuations as expected near a nematic QCP \cite{lederer_enhancement_2015,labat_pairing_2017}, the fall of $T_{c}$ under strain is rationalized as a consequence of the suppression of these fluctuations. 

Within a BCS type of theory~\cite{schrieffer_1964}, with pairing mediated by the 
nematic fluctuations, we expect 
$T_c=\Lambda \, {\rm exp}[-1/(N(0) \chi_{nem})]$~\cite{lederer_enhancement_2015,maier_pairing_2014,labat_pairing_2017,Hirschfeld_2011}, 
where $\Lambda$ is an energy cut off and $N(0)$ is the density of states at the Fermi energy. Thus, we expect the scaling $\delta T_c \propto \delta \chi_{nem}$ (see \cite{SI} for a discussion about the range of validity of this scaling).

To further evaluate this point, we plot on Fig.~\ref{fig3} $\delta\chi_{nem}/\chi_{nem,0}$ as a function of the relative change of $T_c$ under strain, using strain as an implicit parameter. We find that $\frac{\delta T_c}{T_c}\propto\frac{\delta \chi_{nem}}{\chi_{nem,0}}$ over a significant range of strain: the close correlation between the two quantities clearly supports a nematic fluctuations driven SC scenario for Co:Ba122 close to the QCP. We stress that magnetic fluctuations, if anything, are expected to be enhanced by strain \cite{cano_effect_2012} as suggested by the enhanced SDW gap observed at $x=0$ (see Fig. S6 of the Supplemental Materials \cite{SI}) and also by NMR and neutron diffraction measurements \cite{kissikov_uniaxial_2018,tam_uniaxial_2017}. Our elasto-Raman data thus provide a clear distinction between magnetic and nematic fluctuations induced enhancement of $T_c$ in Co:Ba122. We note that in principle the suppression of $\chi_{nem}$ with strain could result from the competition between the superconducting and strain-induced nematic orders. However, because nematic order does not significantly reconstruct the Fermi surface, competition between nematic and SC orders is likely weak although its strength and even its sign may depend on the details of the electronic structure \cite{moon_competition_2012, labat_variation_2020,chen_nematicity_2020}. The weakness of the coupling between both orders is highlighted by the lack of detectable anomaly in the temperature dependence of $\chi_{nem}$ upon crossing $T_c$ as zero strain (see \cite{SI}. The competition scenario is further ruled out by the transport measurements~\cite{malinowski_suppression_2020}, which show that the fall of $T_{c}$ under strain weakens considerably away from the QCP, favoring a fluctuation effect rather than a mere static competition scenario.  

\par
Enhancement of $T_c$ close to a nematic QCP is not universally observed in Fe SC. A particular vexing case is FeSe$_{1-x}$S$_x$ where $T_c$ is even suppressed close to the nematic QCP \cite{reiss_suppression_2017,chibani_lattice_2021}. This was attributed to the coupling to the lattice which cuts off nematic fluctuations, and can significant quench the expected $T_c$ enhancement found in electron-only models \cite{labat_pairing_2017, chibani_lattice_2021,reiss_quenched_2020}. The strength of this coupling is embodied in the nemato-elastic coupling constant $\lambda$ between the structural orthorhombic distortion $\epsilon_{B_{2g}}$ to the nematic order parameter $\phi_{nem}$. Recently, elastocalorimetric measurements by Ikeda et al.~\cite{ikeda_elastocaloric_2020} found that $\lambda$ strongly decreases upon approaching the nematic QCP in Co:Ba122, potentially explaining the observed nematic fluctuations enhanced SC in this compound. The comparison between the two samples in our study can also address this issue since $\lambda$ appears as a prefactor of the relative variation of nematic susceptibility with strain (Eq.~\ref{eqDeltaChiEpsQuadratique}). However, a quantitative comparison of the two suppressions of $\chi_{nem}$ with strain does not support a significant dependence of $\lambda$ with $x$ (see Fig. S8 of the Supplemental Material~\cite{SI}). Given this, the origin of the different behaviors of $T_c$ across the nematic QCP in these two Fe SC materials remains an open question.
\par
In conclusion, our elasto-Raman spectroscopy experiments have revealed a strong correlation between superconducting $T_c$ and nematic fluctuations near the nematic QCP of Co:Ba122. We have shown that combining anisotropic strain with a sensitive probe of the SC state like Raman allows to decouple the different fluctuation channels contributions to SC pairing. We expect that a similar methodology can be employed to reveal nematic fluctuations induced pairing in other materials.
\par
\textit{Acknowledgements} This work is supported by Agence Nationale de la Recherche (ANR grant NEPTUN and ANR grant IFAS). 

\bibliography{biblio}
\newpage

\cleardoublepage
\newpage

\begin{center}
 {\Large{Supplemental Material}}
\end{center}

\setcounter{figure}{0}
\renewcommand{\thefigure}{S\arabic{figure}}
\renewcommand{\thetable}{S\arabic{table}}
\renewcommand{\theequation}{S\arabic{equation}}
\renewcommand{\thesection}{S\arabic{section}}

\subsection*{\label{sec:experimental}Single crystal characterization and details on Raman experiments}

The doping level of the doped compound ($x=0.07$) was determined by wavelength-dispersive spectroscopy (WDS). It was also confirmed by the comparison of the Raman spectra of this sample (without applied strain) to Raman spectra of other samples with doping levels estimated (by WDS) between 6 and 8~\%. The fact that the superconducting peak (centered around 75~cm$^{-1}$) is slightly less pronounced in our $x=0.07$ sample than in less-doped samples confirms that this coumpound is slightly overdoped~\cite{chauviere_impact_2010}.

The samples have dimensions 1.6~mm $\times$ 700 $\upmu$m $\times$ 80 $\upmu$m and 1.5~mm $\times$ 300 $\upmu$m $\times$ 50 $\upmu$m respectively for $x=0$ and $x=0.07$, the stress being applied along the longest dimension. After mounting on the CS130 Razorbill cell, the unstrained suspended length was 1100 and 950 $\upmu$m respectively for $x=0$ and $x=0.07$. Both samples were mounted on the cell with a 50 $\upmu$m thickness of epoxy glue at both ends of the sample, both under and above the sample. We used a Loctite Stycast 2850FT epoxy (with catalyst 24LV) cured at 50\degree C for two hours.
\par
The strain transmission ratio $\epsilon_{x^{\prime}x^{\prime}}/\epsilon_{x^{\prime}x^{\prime}}^{nom}$ was evaluated in a previous work for the parent compound using finite element simulations~\cite{philippe_elasto-raman_2022}: based on the geometrical parameters and the elastic coefficients values: as the crystal strongly softens close to $T_s$, this ratio is close to 1. With the same method, we obtain $\epsilon_{x^{\prime}x^{\prime}} = 0.82 \epsilon_{x^{\prime}x^{\prime}}^{nom}$ for the $x=0.07$ compound below 50~K. This transmission ratio was used to compare our results with ref. \cite{malinowski_suppression_2020} in Figure 3 of the main text.
\par
The Raman scattering geometry was non-collinear, with the incoming photon wave vector arriving at 45\degree ~with respect to the normal of the sample surface plane, and the outgoing photon wave vector lying along the normal. A diode-pumped single longitudinal mode solid-state laser emitting at 532 nm was used for the excitation. The laser spot size is estimated to be less than 50 $\upmu$m, so we expect strain inhomogeneity to have less impact in our Raman measurements than in transport measurements performed in similar cells. The outgoing photons were collected using a X20 long working distance microscope objective (NA=0.28) and analyzed via a triple grating spectrometer equipped with a nitrogen cooled CCD camera. A laser power of 10 mW was used for all spectra for the $x=0$ sample. For the $x=0.07$ sample, the power was 6~mW for all spectra, except at 7~K at which it was 2~mW. Based on previous Raman measurements on Ba122, the laser heating was estimated to be about 1 K/mW \cite{gallais_observation_2013,kretzschmar_critical_2016}. Comparison of spectra for the sample mounted on the uniaxial strain cell with no strain applied, with spectra for unmounted samples, showed that the cantilever geometry does not significantly affect the heating. The temperatures indicated in the main text and in this supplemental material were all corrected for this heating.

\subsection*{\label{sec:computationOP}Nematic order parameter under constant uniaxial strain}

We start from a Landau-Ginzburg expansion of the free energy in a mean-field framework describing
a nemato-structural transition under uniaxial stress. To be consistent with the experimental
conditions and the notations in the main text, the $x$ and $y$ axes are parallel to those of
the the 2-Fe unit cell. In this notation the system experiences a stress $\sigma$ along the
$[110]$ direction, which can be expressed as $\sigma_{x^{\prime}x^{\prime}} = \sigma$, where
$(x^{\prime}, y^{\prime})$ are coordinates $\pi/4$ rotated with respect to $(x, y)$. The stress
tensor $\hat{\sigma}$ in the unprimed coordinates is related to that of the primed coordinates
$\hat{\sigma}^{\prime}$ by $\hat{\sigma} = R^T \hat{\sigma}^{\prime} R$, where
\[
R = \begin{pmatrix}
\cos \theta & \sin \theta \\
- \sin \theta & \cos \theta
\end{pmatrix}
\]
is the rotation matrix and $\theta$ is the angle of rotation. This leads to
$\sigma_{xx} = \sigma_{yy} = \sigma_{xy} = \sigma_{yx} = \sigma/2$. Thus,
in the $(x, y)$ coordinate the induced strain has both
$A_{1g}$ and $B_{2g}$ contributions, while the nematic order parameter has
$B_{2g}$ symmetry. We write the free energy as
\begin{multline}
    F = \frac{a}{2}\phi^{2} + \frac{b}{4}\phi^{4} + \frac{1}{2}C_{A}(\epsilon_{xx}
    + \epsilon_{yy})^{2} \\ + 2C_{66}\epsilon_{xy}^{2}
    - 2\lambda\epsilon_{xy}\phi - \frac{1}{2}\sigma (\epsilon_{xx} + \epsilon_{yy} + 2 \epsilon_{xy})
\label{eq:freeEnergy}
\end{multline}
where $\phi$ is the nematic order parameter, $\epsilon_{xx}$ and $\epsilon_{yy}$ are the longitudinal
strains respectively along the $x$ and the $y$ axes, so that $\epsilon_{xx} + \epsilon_{yy}$ is the
$A_{1g}$ strain with the elastic coefficients $C_A = \frac{1}{2}(C_{11} + C_{12})$, and
$\epsilon_{xy}$ is the $B_{2g}$ shear associated with the elastic constant $C_{66}$.

In the mean-field framework, we have $a = a_0 (T - T_{S}^{0})$, where $a_0$ is a temperature
independent parameter and $T_{S}^{0}$ is the bare nematic transition temperature without
electron-lattice coupling. Indeed, without applied stress, because of the electron-lattice coupling,
the nematic transition is simultaneous to the structural transition at a temperature shifted
from $T_{S}^{0}$ to $T_{S}^{0} + \lambda^{2}/(a_{0}C_{0})$~\cite{bohmer_electronic_2016}.

In our elasto-Raman spectroscopy experiments, it is not the stress $\sigma_{x^{\prime} x^{\prime}}$ but, rather,
the strain $\epsilon_{x^{\prime} x^{\prime}}$ which is a control parameter that is typically held constant
while temperature is varied. So if we define $u =\epsilon_{xx} + \epsilon_{yy}$,
$\psi =2\epsilon_{xy}$ and $\epsilon_{x^{\prime} x^{\prime}} = \epsilon$, then, from the relation
$\epsilon_{x^{\prime} x^{\prime}} = (\epsilon_{xx} + \epsilon_{yy} + 2\epsilon_{xy})/2$ we have the constraint
\begin{equation}
u + \psi = 2\epsilon
\label{eq:constraint}
\end{equation}
with $\epsilon$ a constant. Physically, we are in a regime of constant strain along $x^{\prime}$, which is a
combination of $A_{1g}$ and $B_{2g}$ strains in the $(x, y)$ coordinate,
and the stress $\sigma$ is a free parameter that can adjust itself depending
on the mechanical response of the sample. The values of $\phi$, $u$ and $\psi$ at equilibrium,
which we denote ($\phi_0$, $u_0$, $\psi_0$) are the solutions of the following equations
\begin{equation}
    \frac{\partial F}{\partial \phi} = a\phi + b\phi^3 - \lambda\psi = 0,
    \label{eq:dFdphi}
\end{equation}
\begin{equation}
    \frac{\partial F}{\partial u} = C_A u - \frac{\sigma}{2} = 0,
    \label{eq:dFdu}
\end{equation}
\begin{equation}
    \frac{\partial F}{\partial \psi} = C_{66} \psi - \lambda\phi - \frac{\sigma}{2} = 0.
    \label{eq:dFdpsi}
\end{equation}
Eliminating $(u, \sigma)$ using Eqs.~\eqref{eq:dFdu}, \eqref{eq:dFdpsi} and the constraint, we obtain
%
\begin{equation}
    \psi_0 = \frac{\lambda}{C_A + C_{66}}\phi_0 + \frac{2C_A}{C_A + C_{66}}\epsilon.
    \label{eq:psi0phi0epsilon}
\end{equation}
Next, eliminating $\psi_0$ using~\eqref{eq:dFdphi}, we get
\begin{equation}
    \left(a - \frac{\lambda^2}{C_{66} + C_A}\right)\phi_0 + b\phi_{0}^{3} = \frac{2C_A\lambda}{C_A + C_{66}}\epsilon.
\end{equation}
Thus, $\phi_0$ is given by the solution of
\begin{equation}
    \phi_{0}^{3} + \tilde{a}\phi_{0} - \tilde{\epsilon} = 0,
    \label{eq:equationPhi0}
\end{equation}
where
\begin{equation}
    \tilde{a} = \frac{1}{b}\left(a - \frac{\lambda^2}{C_{66} + C_A}\right),
\end{equation}
\begin{equation}
    \tilde{\epsilon} = \frac{2C_A\lambda \epsilon}{b(C_A + C_{66})}.
\end{equation}

The solution of~\ref{eq:equationPhi0} has three regimes which are given by the following.
\newline
(i) If $\tilde{a} > 0$, then
\begin{equation}
    \phi_{0} = 2 \sqrt{\frac{\tilde{a}}{3}} \sinh\left[\frac{1}{3}\sinh^{-1}\left(\frac{3\tilde{\epsilon}}
    {2\tilde{a}}\sqrt{\frac{3}{\tilde{a}}}\right)\right].
\end{equation}
(ii) If $\tilde{a} < 0$ and $4\tilde{a}^3 + 27\tilde{\epsilon}^2 > 0$, then
\begin{equation}
    \phi_{0} = 2\sqrt{\frac{-\tilde{a}}{3}}\cosh\left[\frac{1}{3}\cosh^{-1}\left(\frac{-3\tilde{\epsilon}}
    {2\tilde{a}}\sqrt{\frac{-3}{\tilde{a}}}\right)\right]  \mathrm{sgn}(\tilde{\epsilon}).
\end{equation}
(iii) If $\tilde{a} < 0$ and $4\tilde{a}^3 + 27\tilde{\epsilon}^2 < 0$, then
\begin{equation}
    \phi_{0} = 2\sqrt{\frac{-\tilde{a}}{3}}\cos\left[\frac{1}{3}\cos^{-1}\left(\frac{-3\tilde{\epsilon}}
    {2\tilde{a}}\sqrt{\frac{-3}{\tilde{a}}}\right)\right].
\end{equation}
Note that in regimes (i) and (ii), $\phi_0$ is odd in $\epsilon$, while in regime (iii),
$\phi_0(\epsilon) = \phi_0(0)+ \delta \phi_0(\epsilon)$ and $\phi_0$ has no particular parity in general.

\subsection*{\label{sec:computationChi}Thermodynamic nematic susceptibility}

We introduce a nematic force $s$ which appears in the free energy expression~\eqref{eq:freeEnergy}
as an additional term $-s\phi$, such that \eqref{eq:dFdphi} changes to
\begin{equation}
    a\phi + b\phi^3 - \lambda\psi = s,
    \label{eq:dFdphiS}
\end{equation}
while \eqref{eq:dFdu} and \eqref{eq:dFdpsi} remain the same. Let the solutions of \eqref{eq:dFdu},
\eqref{eq:dFdpsi} and \eqref{eq:dFdphiS} be ($\phi_0 + \delta\phi$, $u_0 + \delta u$, $\psi_0 + \delta\psi$),
with the shifts ($\delta\phi$, $\delta u$, $\delta\psi$) due to the introduction of $s$. Then, to
linear order in the shifts, we get
\begin{equation}
    a\delta\phi + 3b\phi_0^2\delta\phi - \lambda\delta\psi = s.
\end{equation}

Since $\epsilon$ is a constant, from \eqref{eq:psi0phi0epsilon}, we get
\begin{equation}
    \delta\psi = \frac{\lambda}{C_A + C_{66}} \delta\phi,
\end{equation}
which implies
\begin{equation}
    \left(a - \frac{\lambda^2}{C_A + C_{66}} + 3b\phi_0^{2}\right)\delta\phi = s.
\end{equation}
Thus, the thermodynamic nematic susceptibility $\chi_{nem,th}$ under constant strain is given as:
\begin{equation}
    \chi_{nem,th}^{-1} = a - \frac{\lambda^2}{C_A + C_{66}} + 3b\phi_0^{2}
    \label{eq:thermoChiNem}
\end{equation}
where $\phi_0$ is given by the solution of \eqref{eq:equationPhi0}.

Note that, for a spontaneous nemato-structural transition without applied stress the coupling of
the electrons with the lattice shifts the bare thermodynamic nematic susceptibility from
$\chi_{nem,th,0}^{-1} = a$ to $\chi_{nem,th}^{-1} = (a - \lambda^2/C_{66})$. Here, since the experiment
is at constant applied $\epsilon$, the shift due to the lattice is not $- \lambda^2/C_{66}$
but $- \lambda^2/(C_{66} + C_A)$.

Also, we can write $a - \lambda^2/(C_A + C_{66}) = a_0(T - T_1)$, so that
\begin{equation}
    T_1 = T_s^0 + \frac{\lambda^2}{a_0(C_A + C_{66})}.
\end{equation}
This implies that $T_s^0 < T_1 < T_s$. Note that regime (i) defined above corresponds to $T > T_1$.
In practice, the difference between $T_s$ and $T_1$ can be small, and they remain unresolved in our experiment.

\subsection*{\label{sec:computationChiDynamic}Dynamical nematic susceptibility}

\begin{figure}
    \includegraphics[width=0.49\textwidth]{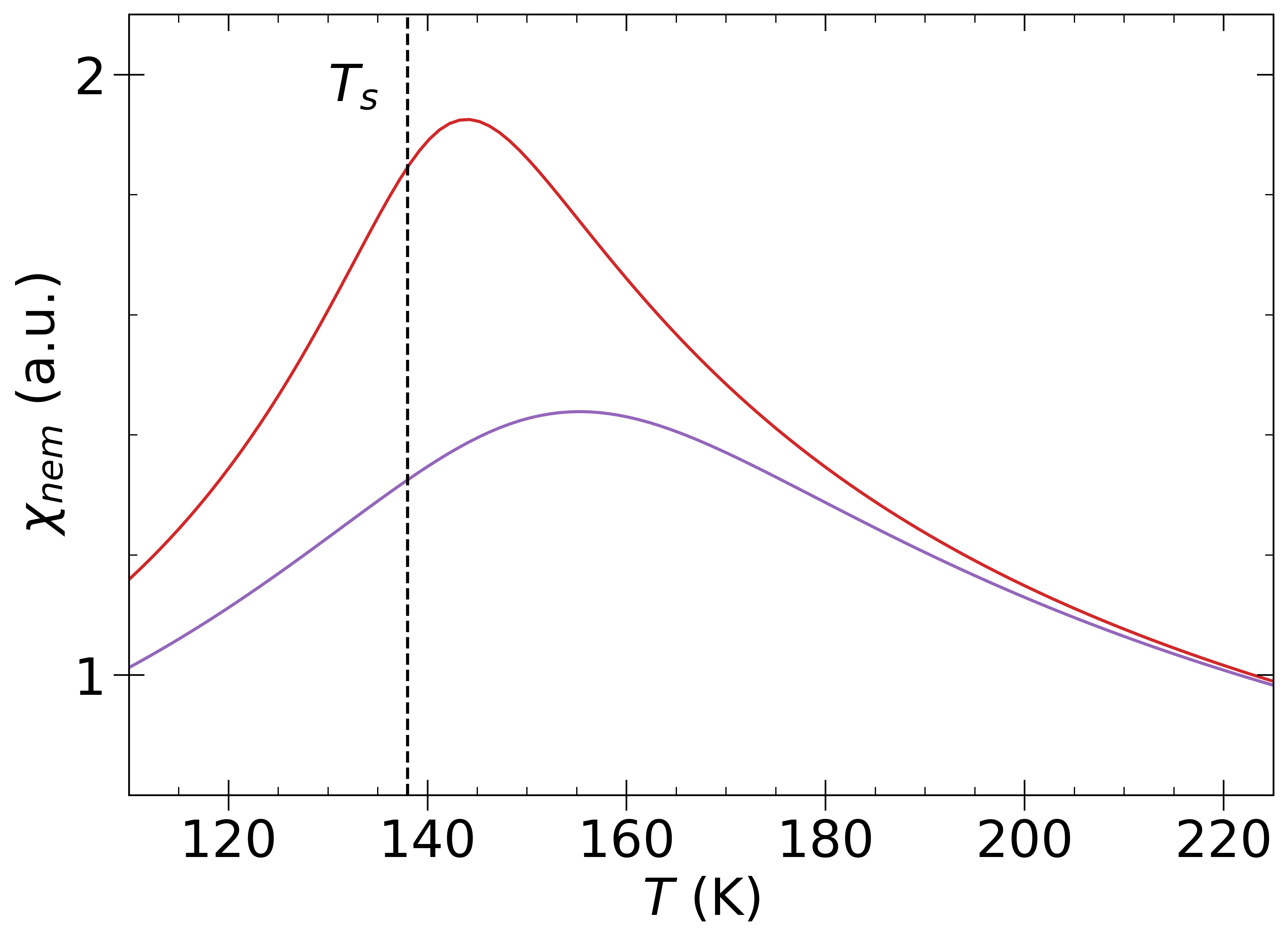}
    \caption{Theoretical computation of the dynamical nematic susceptibility $\chi_{nem}(T, \epsilon)$
    as a function of temperature $T$ for low (red) and high (purple) uniaxial strains
    $\epsilon = -(1.5, 6.5) \times 10^{-3}$, respectively, obtained using Eq.~\eqref{eq:chiNemStaticRaman},
    with $\phi_0$ solution of Eq.~\eqref{eq:equationPhi0}. The dashed vertical line indicates the nematic transition temperature $T_s = 138$~K in the absence of stress. See text for the values
    of the various parameters used here, chosen to qualitatively fit the experimental data in Fig. 1(h) of the main text. Increasing uniaxial strain reduces the nematic correlation
    length and, therefore, flattens the peak in the $T$ dependence. It also pushes the maximum to higher temperatures.}
    \label{figS0}
\end{figure}

The nematic susceptibility measured by the Raman response is different from the thermodynamic
one computed previously and given in~\eqref{eq:thermoChiNem}. To get the Raman nematic susceptibility
we need to look at fluctuations at finite ($q$, $\omega$), and then take the dynamical limit
$\frac{q}{|\omega|} \rightarrow 0$~\cite{gallais_charge_2016}. The nematic susceptibility in the
dynamical limit ($|\omega_n| \gg v_{F} q$, with $v_F$ the Fermi velocity) is~\cite{gallais_charge_2016}:
\begin{multline}
    \chi_{nem,dyn}^{-1} (q, i\omega_n)= a + 3b\phi_0^{2} \\ + c_1 \left(\frac{q}{k_F}\right)^2
    + c_2\frac{|\omega_n|}{|\omega_n| + \tau^{-1}},
    \label{eq:dynChiNem}
\end{multline}
where $k_F$ is the Fermi wave vector, $\tau$ is the electron lifetime from impurity scattering, and the
coefficients $c_{1,2}$ depend on the details of the band structure. Then, by the Kramers-Kronig relation,
the static nematic susceptibility $\chi_{nem}$ obtained in Raman spectroscopy as the area under the
Raman conductivity (Raman response divided by the Raman shift) is
\begin{equation}
    \chi_{nem}^{-1} \equiv \lim_{\omega_n \to 0} \chi_{nem,dyn}^{-1}(0,\omega_n) = a + 3 b \phi_0^2.
    \label{eq:chiNemStaticRaman}
\end{equation}
Thus, the term $\lambda^2/(C_A + C_{66})$ that enters the thermodynamic
nematic susceptibility $\chi_{nem, th}$~\eqref{eq:thermoChiNem} is not present
in the Raman nematic susceptibility $\chi_{nem}$.
Note, however, that $\phi_0$ is given by the
solution of~\eqref{eq:equationPhi0} and will follow thermodynamics.

In Fig.~\ref{figS0} we show the theoretical temperature dependence of $\chi_{nem}(T, \epsilon)$
for two different fixed uniaxial strains. The parameters were chosen to qualitatively fit the experimental data (Fig. 1(h) of the main text). We used: $a_0 = 0.05/$ (\AA)$^{3}$, $b = 40$ K/(\AA)$^{3}$, $C_{11} =$
93 GPa, $C_{12} =$ 17 GPa, $C_{66} = $ 30~GPa, $\lambda =$ 0.135 $C_{11}$, $T_s^0 =$ 100~K. The elastic coefficients values are experimental ones obtained in~\cite{fujii_diverse_2018}, and $T_s^{0}$ is experimentally assessed through Raman spectroscopy in~\cite{gallais_observation_2013}. Note that to ensure $T_s = $ 138~K at zero strain, $a_0$, $\lambda$ and $b$ can not be chosen independently. To qualitatively fit the experimental data, we added to $\chi_{nem}$ a strain and temperature independent constant (taken here as 0.35 a.u.).

At low uniaxial strain there is a
well-defined peak in the temperature dependence of $\chi_{nem}$, which is due to the presence of nematic
fluctuations that would be critical in the absence of uniaxial strain. The role of the uniaxial strain is
to reduce the nematic correlation length and, therefore, to broaden the peak, as seen for high values of the strain. In the absence of $\epsilon$ the peak is expected at $T_1$, which then gets shifted to higher temperatures as $\epsilon$ increases. This theoretical behavior captures qualitatively the trend seen in the experiments and reported in Fig.~1(h) of the main text.

We introduce $\delta\chi_{nem} = \chi_{nem}(\epsilon) - \chi_{nem,0}$ with
$\chi_{nem,0} = \chi_{nem}(\epsilon = 0)$, and obtain
\begin{equation}
\label{eq:S22}
    \frac{\delta\chi_{nem}}{\chi_{nem,0}} = 3b[\phi_0^2(\epsilon=0) - \phi_0^2(\epsilon)]
    \chi_{nem}(\epsilon).
\end{equation}
It is this quantity which has been plotted in Figs.~ 1(g), 2(e) and 3 of the main text.

Next, we restrict ourselves to the case $\phi_0^2(\epsilon=0) = 0$ (regimes (i) and (ii) for the
solution of~\eqref{eq:equationPhi0}), which is valid in particular for $T > T_s$. In the low strain
regime, where we can neglect the nonlinear term in Eq.~\eqref{eq:equationPhi0}, we get
\begin{equation}
 \phi_0 \approx \frac{2 C_A \lambda \epsilon}{a(C_A + C_{66}) - \lambda^2}.
\end{equation}
%

By symmetry, $\delta \chi_{nem}$ is an even function of the applied
strain $\epsilon$. To quadratic order accuracy, we can replace 
$\chi_{nem}(\epsilon)$ in Eq.~(\ref{eq:S22}) by $\chi_{nem,0}$. This implies
\begin{equation}
\frac{\delta\chi_{nem}}{\chi_{nem,0}} \approx
- \frac{12 b \lambda^2 \epsilon^2 C_A^2}{a [a(C_A + C_{66}) - \lambda^2]^2}.
\end{equation}
Using $\chi_{nem,0}=\frac{1}{a}$ we obtain:
\begin{equation}
 \frac{\delta\chi_{nem}}{\chi_{nem,0}} \approx
- 12 b \lambda^2\chi_{nem,0}^3 \epsilon^2\frac{ C_A^2}{(C_A+\tilde{C}_{66})^2}   
\end{equation}
where $\tilde{C}_{66}$ is the shear modulus renormalized by nematic flcutuations: $\tilde{C}_{66}=C_{66}-\lambda^2\chi_{nem,0}$
In other words $\delta\chi_{nem} \propto \epsilon^2$ in the low strain regime with a slope that strongly grows with the zero-strain nematic susceptibility $\chi_{nem,0}$.

\subsection*{\label{sec:nematicResonance}Nematic resonance in the superconducting state under strain}

In a fully gapped superconducting state, in the presence of nematic fluctuations, the usual pair
breaking peak seen in Raman response at frequency $2\Delta$ gets transformed into a resonance feature
at frequency $\Omega_r < 2\Delta$. Here, we first briefly review the theory of the
resonance~\cite{gallais_nematic_2016}, and then we consider the effect of strain.

For a fully gapped superconductor the frequency dependence of the nematic susceptibility is given
by~\cite{gallais_nematic_2016}
\begin{equation}
\label{eq:chi_nem_sc}
\chi_{nem,dyn}(\omega) = \frac{\chi_0(\omega)}{1 - g \chi_0(\omega)},
\end{equation}
where $g > 0$ is an interaction parameter that triggers the nematic transition
in the appropriate regions of the phase diagram, and whose microscopic origin is not
relevant for the current discussion. For systems in the clean limit $\Delta > 1/\tau$,
where $\tau$ is the impurity scattering induced electron lifetime, the bare nematic polarization
is given by~\cite{gallais_nematic_2016}
\begin{eqnarray}
\chi_0(\omega + i\eta) &=& \rho \frac{\arcsin (\bar{\omega})}{\bar{\omega}
\sqrt{1 - \bar{\omega}^2}}, \quad \quad \bar{\omega} \leq 1,
\nonumber \\
&=&
\rho \frac{- {\rm arcosh}(\bar{\omega}) + i \pi/2}{\bar{\omega} \sqrt{\bar{\omega}^2 -1}},
\quad \quad \bar{\omega} > 1.
\nonumber
\end{eqnarray}
where $\bar{\omega} \equiv \omega/(2\Delta)$, and $\rho \equiv \sum_{\bf k}
\gamma_{B_{2g}}({\bf k})^2 \delta(\epsilon_{\bf k} - \epsilon_F)$ is the $B_{2g}$ density of states,
$\epsilon_{\bf k}$ is the electron dispersion and $\gamma_{B_{2g}}({\bf k})$ is a form factor that
transforms as $k_x k_y$.

%
Note that, since $\rho$ is independent of the superconducting gap $\Delta$,
the zero frequency limit of the dynamical susceptibility $\chi_{nem,dyn}(\omega \rightarrow 0)$ is
also independent of $\Delta$~\cite{labat_variation_2020}.

The nematic resonance occurs at a frequency $\Omega_r$ where the real part of the denominator
in Eq.~\eqref{eq:chi_nem_sc} vanishes, i.e.:
\begin{equation}
1 - g \rho f(\bar{\omega}_r) = 0
\label{eq:resonanceFrequencyDef}
\end{equation}
where $\bar{\omega}_r = \Omega_r / (2\Delta)$, and $f(\bar{\omega}) = \arcsin(\bar{\omega})/(\bar{\omega}
\sqrt{1 - \bar{\omega}^{2}})$. Since we consider the ground state of the system to be non-nematic, we
have $g \rho < 1$. On the other hand, the function $f(\bar{\omega})$ diverges at $\bar{\omega} =1$.
These two limiting behavior guarantees the existence of the resonance at a frequency $\Omega_r < 2\Delta$.
In the vicinity of the resonance the dynamical nematic susceptibility is given by
$\chi_{nem,dyn}(\omega) \sim Z \delta(\omega - \Omega_r)$, where the weight of the resonance is given by
\begin{equation}
 \bar{z} = \frac{2\pi}{g}  \zeta({\bar{\omega}_r})
   \label{eq:resonanceWeightDef}
\end{equation}
where $\bar{z} \equiv Z/\Delta$, and $\zeta(x) = f(x) / f^{\prime}(x)$. The advantage of
working with the spectral weight scaled by the gap value will become evident below.

We consider now the effect of strain. We assume that the interaction $g$ is unchanged by strain.
Thus, the main effect of strain is on the electronic dispersion and, hence, on $\rho$. We consider
the system is in a phase where $\phi_0 =0$ in the absence of strain. Then, by making contact with
the phenomenological Landau-Ginzburg theory we get that, in the absence of strain
\[
\chi_{nem,0}^{-1} = 1/\rho - g = a,
\]
where $\chi_{nem,0}$ is the value of the nematic susceptibility without strain, and the
Landau-Ginzburg parameter $a$ is defined through Eq.~\eqref{eq:freeEnergy}. While, in the
presence of strain
\[
\chi_{nem}(\epsilon)^{-1} = 1/(\rho + \delta \rho) - g = a + 3b \phi_0^2.
\]
Comparing the two equations we get
\begin{equation}
\label{eq:delta-chi-nem}
\frac{\delta \chi_{nem}(\epsilon)}{\chi_{nem}(\epsilon)} = - 3 b \phi_0^2 \chi_{nem,0},
\end{equation}
and
\begin{equation}
\delta\rho =  -3b\phi_{0}^{2}\rho^2.
\end{equation}

Our next step is to deduce how the change $\delta \rho$ affects the nematic resonance. For simplicity,
in the following we assume that $\Delta$ is unchanged by strain. Later we comment on effects of the
change in $\Delta$ due to the strain. By differentiating~\eqref{eq:resonanceFrequencyDef}, the shift
in the resonance frequency can be expressed as
\begin{equation}
\label{eq:delta-omega-r}
 \delta\bar{\omega}_r = 3b\phi_{0}^{2}\rho\zeta(\bar{\omega}_r).
\end{equation}
\begin{figure*}
    \centering
    \includegraphics[width=0.98\textwidth]{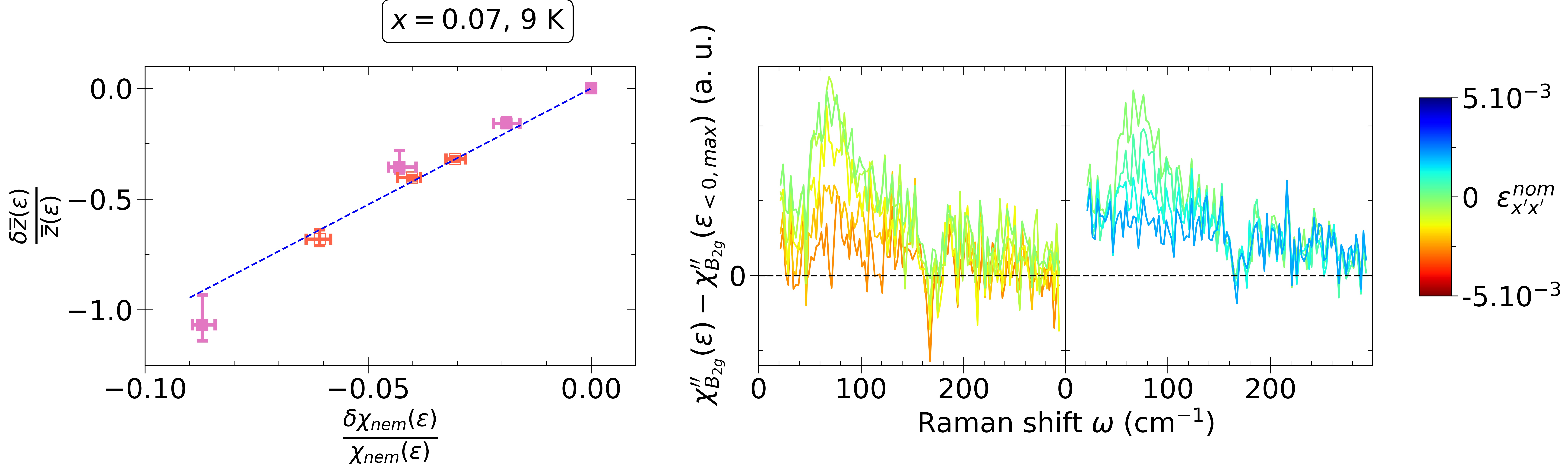}
    \caption{Scaling between $\chi_{nem}$ and the nematic resonance spectral weight divided by $T_c$: $\tilde{z}=\frac{z}{T_c}$ (pink / orange data points correspond to compressive / tensile strains. To extract the spectral weight $z$ of the SC peak, the spectra at high strain were used as a reference and subtracted from the spectra at low strain (right). The resulting difference spectra were then integrated to obtain $z$ which was then divided by the corresponding $T_c$ to obtain $\tilde{z}$.}
    \label{figRes}
\end{figure*}

This function diverges at $x\rightarrow 0$, decreases to 0
at $x=1$, and stays positive between 0 and 1. Since $\delta\bar{\omega}_r > 0$, we conclude that,
with strain the resonance mode moves towards higher energy approaching $2\Delta$. Furthermore, the strain induced change in the weight of the resonance can be obtained by differentiating~\eqref{eq:resonanceWeightDef}. We obtain
\begin{equation}
    \delta \bar{z} = \frac{2\pi}{g}  \zeta^{\prime}(\bar{\omega}_r) \delta\bar{\omega}_r.
    \label{eq:delta-Z}
\end{equation}
$\zeta^{\prime}(x)$ is negative
in the range $x = (0, 1)$, which implies $\delta \bar{z} < 0$. The weight (so the
height and width) of the resonance mode decreases upon strain, which is consistent with our experimental
observation. Finally, comparing the above equations we get the scaling relation
\begin{equation}
\frac{\delta \bar{z} (\epsilon)}{ \bar{z} (\epsilon)} =
\kappa \frac{\delta \chi_{nem}(\epsilon)}{\chi_{nem}(\epsilon)},
\end{equation}
where $\kappa = -\rho \zeta^{\prime}(\bar{\omega}_r)/\chi_{nem,0}$ is a constant independent of
strain.

Note, in the preceding lines we assumed that $\Delta$ is constant with strain. In reality, around the nematic
quantum critical point $T_c$ decreases strongly with strain, indicating that $\Delta$ decreases also with
strain. However, since Eqs.~\eqref{eq:resonanceFrequencyDef} and \eqref{eq:resonanceWeightDef}
involve scaling functions of
$\omega/(2\Delta)$, the effect of the change in $\Delta$ is already included in these equations.
In practice, we do not have access to the gap value $\Delta$, and so we scale $Z$ by $T_c$ to get
$\bar{z}$, which assumes that $\Delta \propto T_c$. Note also, that Eqs. \ref{eq:delta-omega-r} and \ref{eq:delta-Z} implicitly assume that, with application of strain, the variations are linearly related. Thus, the above scaling relation is valid only in the low strain regime where nonlinear effects can be neglected. In figure \ref{figRes} we show that the above scaling relation between $\chi_{nem}$ and $\tilde{z}$ is satisfied at low strain in the $x$=0.07 crystal. 

\subsection*{Nematic fluctuation mediated superconductivity}

From a theory point of view, this topic has been discussed in the recent 
literature, see, e.g. Refs.~\cite{lederer_enhancement_2015,labat_pairing_2017}. In these scenarii the SC pairing of Fe SC is mediated via two channels: the first one is non-critical and might be of magnetic origin, while the second one is nematic. The latter is relevant only close to the QCP where the nematic fluctuations are expected to contribute to a significant fraction of the pairing strength.
Within a Bardeen-Cooper-Schrieffer theory, the superconducting transition temperature is given by $T_c = \Lambda \, {\rm exp}[-1/(N(0) V)]$, where $N(0)$ is the density of states at the Fermi level, $\Lambda$ a cut-off energy scale, and $V$ is the pairing
interaction.

In general $V$ will have contributions coming from critical nematic fluctuations and non-critical (likely magnetic) fluctuations. $V$=$\chi_{nem}+\chi_{NC}$ where $\chi$ are electronic susceptibilities in the corresponding channel in units of energy. Assuming anisotropic strain only influence $\chi_{nem}$ we can write $\chi_{nem}=\chi_{nem,0}+\delta\chi_{nem}$ where $\chi_{nem,0}$ is the nematic susceptibility at zero strain. Close to the QCP the SC pairing is dominated by the nematic fluctuations mediated pairing channel, and $V$ can be replaced by the nematic susceptibility $\chi_{nem}$ \cite{lederer_enhancement_2015}. Neglecting non-critical fluctuations $\chi_{NC}=0$ and assuming $\frac{\delta\chi_{nem}}{\chi_{nem,0}}$ is small we can linearize:
\begin{equation}
\frac{\delta T_c}{T_c}=\frac{1}{\chi_{nem,0}N_0}\times\frac{\delta\chi_{nem}}{\chi_{nem,0}}
\end{equation}
This is the scaling found in fig. 3 of the manuscript. We note that since $\frac{\delta\chi_{nem}}{\chi_{nem,0}}$ reaches up to 0.2 in the experiment one can question the validity of the above linearization for such large $\delta \chi_{nem}$. However, note that experimentally we use a finite high energy cut-off to extract $\chi^{exp}_{nem}$. Therefore the absolute value of $\chi_{nem}$ is underestimated, i.e. $\chi^{exp}_{nem}<\chi_{nem}$. Contrary to $\frac{\delta T_c}{T_c}$ the ratio $\frac{\delta\chi_{nem}}{\chi_{nem,0}}$ is thus only known up to a multiplication factor which is less than 1 since the energy range where $\chi''_{B_{2g}}(\omega)$ changes is below the cut-off energy.  Given the flat Raman continuum extending up to at least 1000 cm$^{-1}$ (as observed for BaFe$_2$As$_2$) we can estimate $\chi_{nem}$  to be at least 50\% higher than $\chi^{exp}_{nem}$. In addition non-critical fluctuations which contribute to $\chi_{NC}$ will further affect the above approximation. To further evaluate the validity of the linearization in the more general case, we compare below the data with the theoretical expectation without relying on linearization. In the more general case we have:
\begin{equation}
\delta T_c = T_c-T_c^0=\Lambda(e^{-\frac{1}{N_0(\chi_{nem,0}+\delta\chi_{nem}+\chi_{NC})}}-e^{-\frac{1}{N_0(\chi_{nem,0}+\chi_{NC})}})
\end{equation}
\begin{equation}
\frac{\delta T_c}{T_c} =\frac{e^{-\frac{1}{N_0(\chi_{nem,0}+\delta\chi_{nem}+\chi_{NC})}}-e^{-\frac{1}{N_0(\chi_{nem,0}+\chi_{NC})}}}{e^\frac{1}{N_0(\chi_{nem,0}+\chi_{NC})}}
\end{equation}
We set $\chi_{nem,0}=1$, keep $N_0$ free and compare the equation above with our data in 4 scenarii. (1) $\chi_{NC}=0$ and $\chi_{nem,0}^{exp}=\chi_{nem,0}$, (2) $\chi_{NC}=0.3\chi_{nem,0}$ and  $\chi_{nem,0}^{exp}=\chi_{nem,0}$, (3) $\chi_{NC}=0$ and  $\chi_{nem,0}^{exp}=\frac{2}{3}\chi_{nem,0}$. (4) $\chi_{NC}=0.3\chi_{nem,0}$ and  $\chi_{nem,0}^{exp}=\frac{2}{3}\chi_{nem,0}$. We note $\chi_{NC}$=0.3 is likely a lower bound for non-critical fluctuations given the sizable $T_c$ observed in Co-Ba122 even far away from the QCP. Larger values for $\chi_{NC}$ will further extend the approximate linear regime.
While a clear deviation for the last data point is observed for (1), adding non-critical fluctuation and/or taking into account the underestimation of $\chi_{nem,0}$ both give good agreement with the data using  $N_0\approx (0.25 - 0.5) \chi_{nem,0}$.

\begin{figure}
\includegraphics[width=9.5cm]{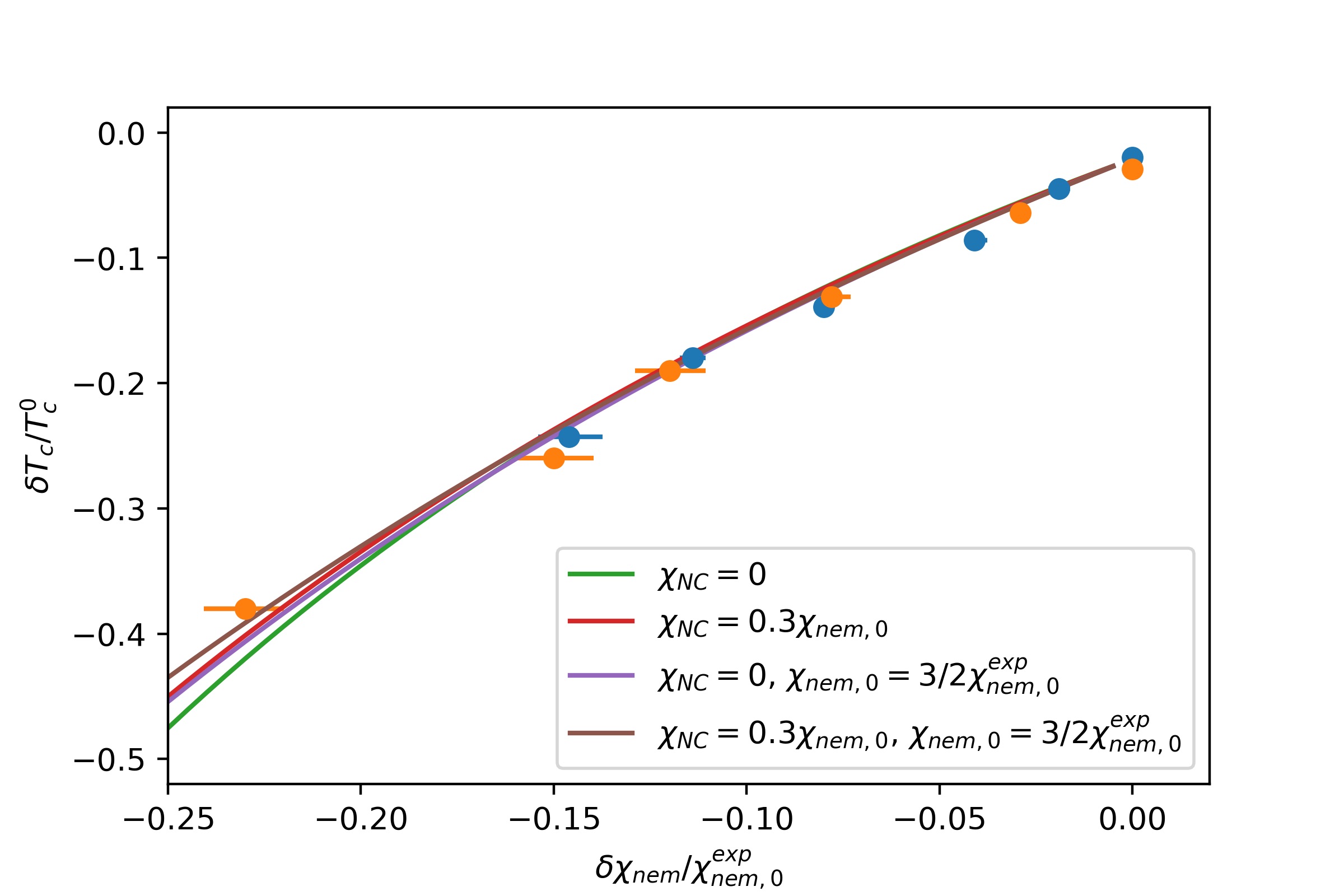}
\caption{Comparison between data under compression (orange points: 9K, blue points: 26K) and theoretical expectations based on the BCS formula using 4 different set of parameters.}
\end{figure}

\subsection*{\label{sec:B2g-Tc} Nematic susceptibility across $T_c$ at zero strain}
\begin{figure*}
    \centering
    \includegraphics[width=0.70\textwidth]{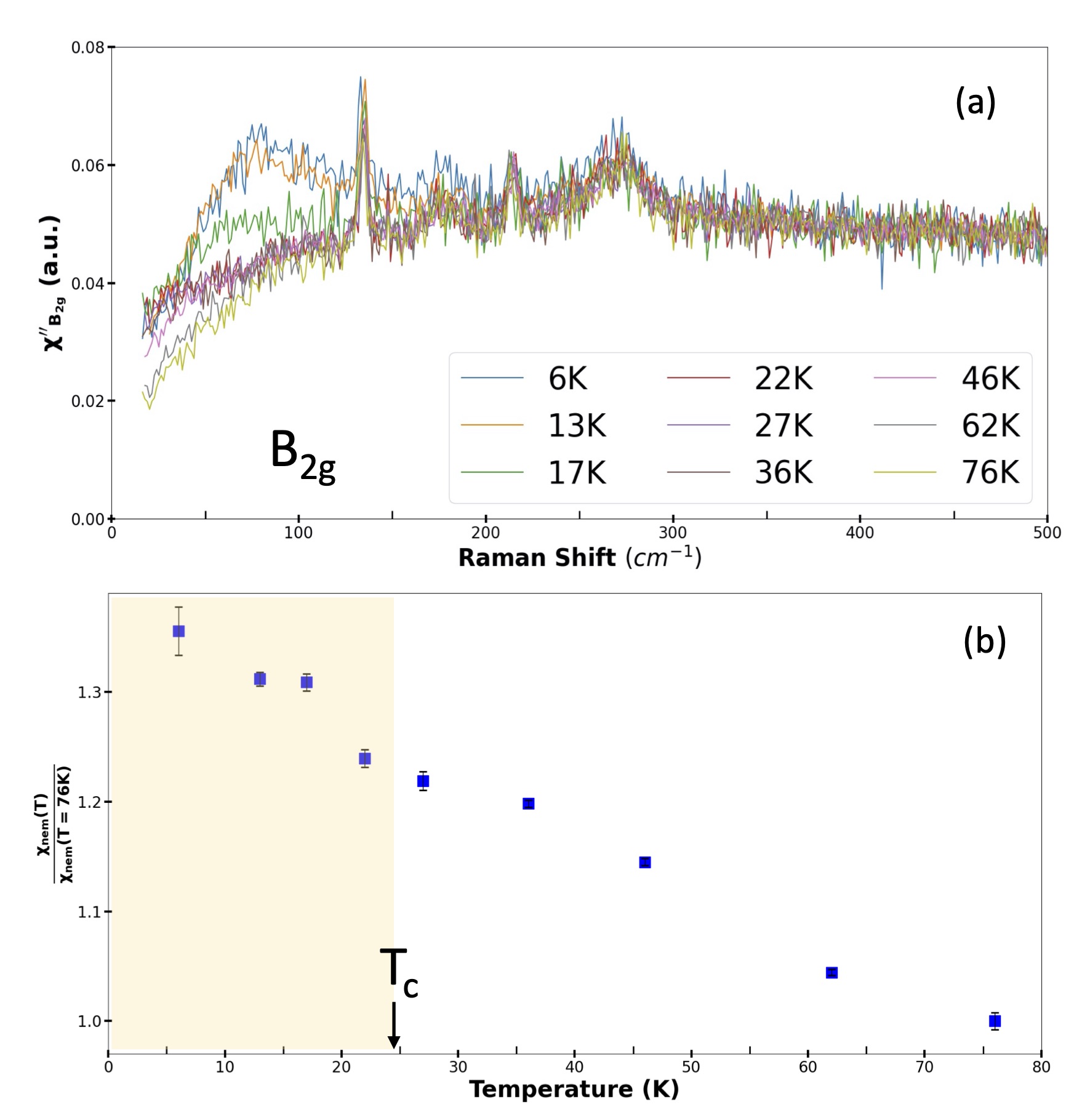}
    \caption{(a) $B_{2g}$ Raman response as a function of temperature at zero strain for the $x=0.07$ crystal. (b) Extracted nematic susceptibility as a function of temperature. The error bars correspond to uncertainties associated to different choices of interpolation schemes for the Raman response down to zero energy. Additional uncertainties related to small changes in spot location and/or optical alignment upon varying temperature are not taken into account and likely dominate.}
    \label{figchiTc}
\end{figure*}

In figure \ref{figchiTc} we show the Raman response and calculated nematic susceptibility as a function of temperature for the $x=0.07$ crystal. Within our accuracy there is no detectable change in $\chi_{nem}$ accross $T_c$ indicating weak competition between superconductivity and nematic order.

\subsection*{\label{sec:B1g}Strain effect on the Raman response in the $B_{1g}$ channel}

As Raman spectroscopy is a symmetry-resolved probe, the nematic susceptibility is measured only through spectra probing the nematic symmetry, that is for Co:Ba122 the $B_{2g}$ symmetry. We probe this nematic channel by choosing the polarizations of the incident and scattered lights respectively along the [100] and [010] directions of the 2-Fe unit cell, and the resulting spectra and analysis are presented in the main text. In this section, we present the effect of strain on the Raman response $\chi^{\prime\prime}_{x^{\prime}y^{\prime}}(\omega)$ in the $B_{1g}$ symmetry, when the polarizations of the incident and scattered lights are respectively along the [110] and [1$\bar{1}$0] directions of the 2-Fe unit cell (so along the $x^{\prime}$ and $y^{\prime}$ axes of the rotated frame we use to describe our experiments). From previous experiments~\cite{gallais_observation_2013}, it is known that, in the parent compound and in this symmetry, the Raman spectra and so the dynamical Raman susceptibility change hardly across $T_s$, showing that the susceptibility is featurless in this transverse symmetry.

\begin{figure*}
    \centering
    \includegraphics[width=0.80\textwidth]{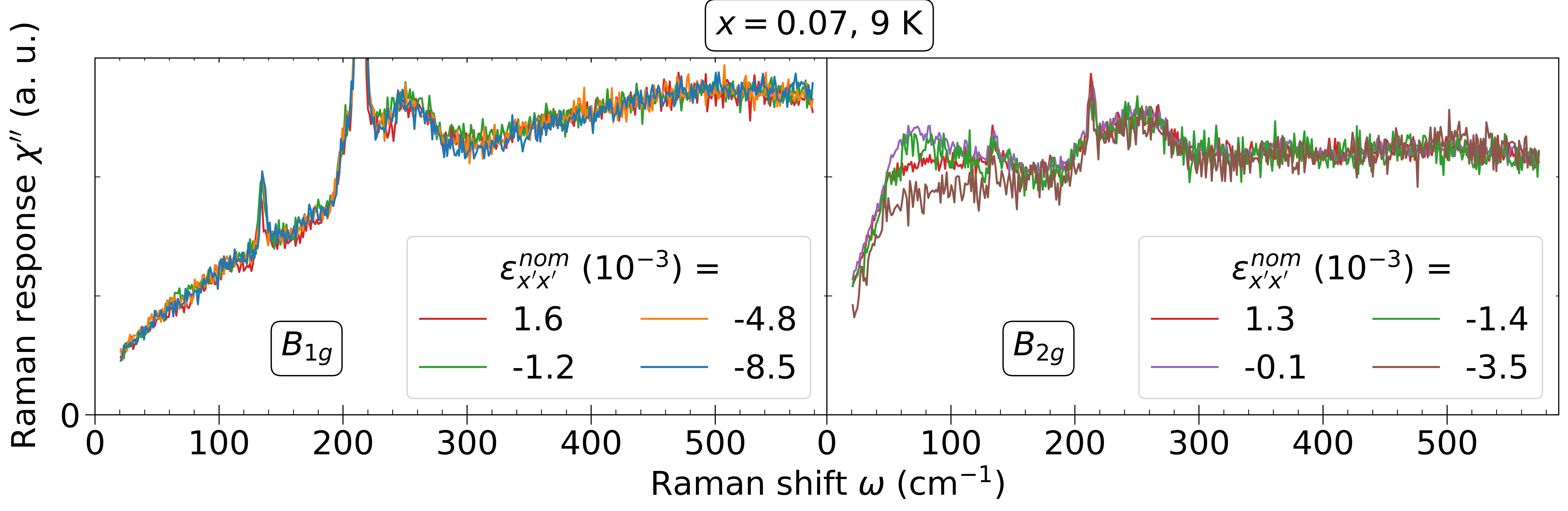}
    \caption{Comparison of the effect of strain between the $B_{1g}$ Raman response (left panel) and the $B_{2g}$ one (right panel) in the $x=0.07$ compound at 9~K. The $B_{2g}$ spectra displayed here are the same as those displayed in Fig. 2 of the main text.}
    \label{figB1g}
\end{figure*}

On Figure~\ref{figB1g}, we compare the effect of strain on the $x=0.07$ compound at low temperature between the $B_{1g}$ and the $B_{2g}$ Raman responses. Across a wide range of mostly compressive strain, we observe no clear variation of the $B_{1g}$ Raman spectra comparable to the one occurring on the same sample in the $B_{2g}$ channel. This observation confirms our interpretation of the effect of strain in the nematic channel in terms of a decrease of the nematic fluctuations upon strain.

\subsection*{\label{sec:deviationDrude}Low energy extrapolation and strain-induced deviation from the Drude lineshape}

To compute the nematic susceptibility (Eq. (1) of the main text), we need to extrapolate the experimental spectra from the low frequency cut-off (around 15-20~cm$^{-1}$ depending on the spectra) down to zero energy. 

In the superconducting state, the shape of the Raman response depends strongly on the anisotropy of the superconducting gap. We extrapolated the spectra using a simple linear form, which seems appropriate for Co:Ba122, in which the gap is believed to display an anisotropic $s^{*}$-type shape \cite{chauviere_impact_2010}. 

In the metallic state, the standard Raman response follows a Drude-like expression at low energy, given by:
\begin{equation}
    \chi^{\prime\prime} (\omega) = \chi_{0}\frac{\Gamma\omega}{\Gamma^2 + \omega^2}
\end{equation}
with $\Gamma$ the relaxation rate of the charge carriers, and $\chi_{0}$ the static Raman response. In the presence of a nematic instability, this Drude form displays a strong temperature dependence close to $T_s$, through both $\Gamma$ and $\chi_{0}$~\cite{gallais_charge_2016}. From previous works, the zero strain $B_{2g}$ Raman response is well fitted with such expression, at least at low energy below 150-200~cm$^{-1}$~\cite{gallais_observation_2013}. 

\begin{figure*}
    \centering
    \includegraphics[width=0.90\textwidth]{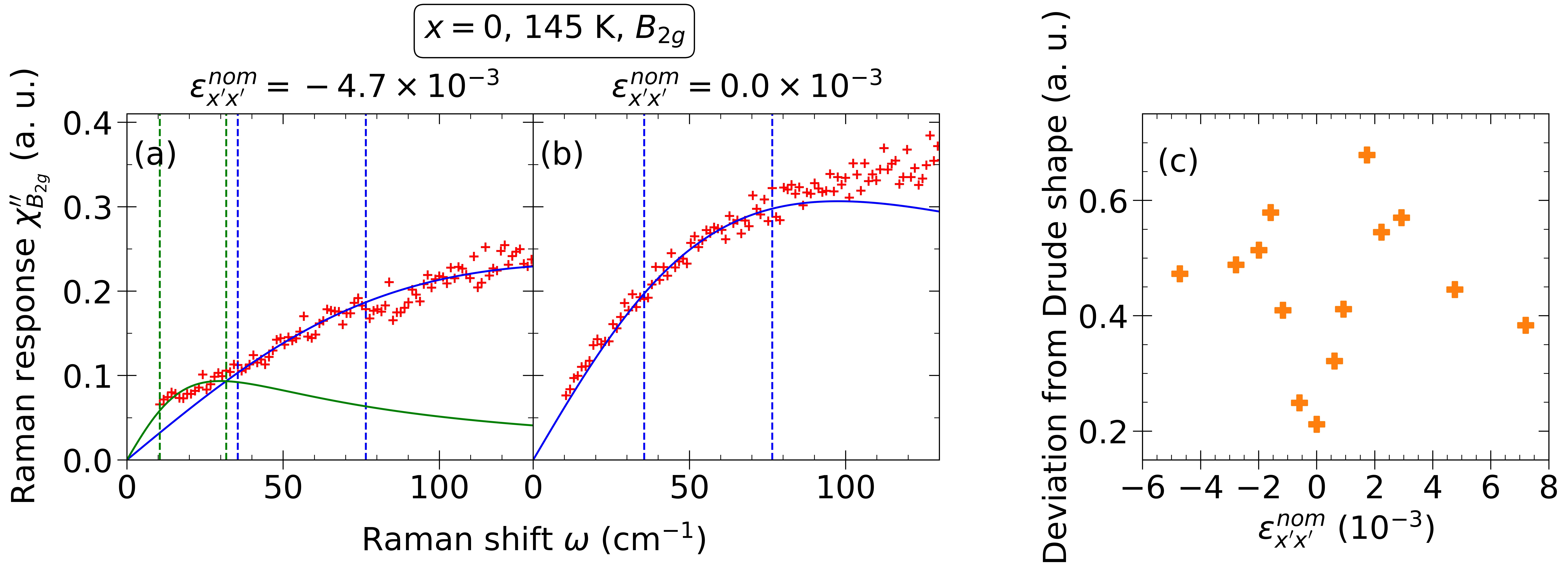}
    \caption{Strain-induced deviation from the Drude shape at low energy for the nematic ($B_{2g}$) Raman response of the parent compound at 145~K. (a) and (b) Fittings of the Raman response at low energy. The red points denote the experimental spectra. Considering a fitting area between 35 and 75~cm$^{-1}$ (between the two blue vertical dashed lines), the zero-strain spectrum (b) is well fitted with a Drude form down to the lowest frequency, whereas the strong-strain one (a) displays a clear deviation (blue curves). Considering a fitting area between 15 and 30~cm$^{-1}$ (between the two green vertical dashed lines) to fit the strong-strain spectrum, the fitting is irrelevant at higher energy above 30~cm$^{-1}$ (green curve). (c) Quantitative assessment of the deviation. The quantity plotted here (deviation from the Drude shape) is defined as the area between the Drude fitting (considering a fitting reference region between 35 and 75~cm$^{-1}$) and the experimental data points, for $\omega < 35$~cm$^{-1}$.}
    \label{figDrudeDeviation}
\end{figure*}

However, for the $x = 0$ sample, we observed a strain-induced deviation from the Drude lineshape for our spectra close to $T_s$, at 145~K. This phenomenon is depicted in Fig~\ref{figDrudeDeviation}, with a tentative quantitative assessment. The deviation is not observed at higher temperature (188~K), but below $T_s$ at 118~K, we observe a significant deviation at all strains. For the $x=0.07$ compound in the metallic state, we do not observe any clear deviation from the Drude shape even under strain. The fact that this deviation from the Drude shape is seen below $T_s$ and above and close to $T_s$ under strain, and not in the $x=0.07$ compound, indicates an intrinsic origin, likely related to a strong static nematic order, but we lack theoretical insight to conclude more deeply about this phenomenon.  

\subsection*{\label{sec:magnetism}Strain effect on the SDW gap}

\begin{figure*}
    \centering
    \includegraphics[width=0.99\textwidth]{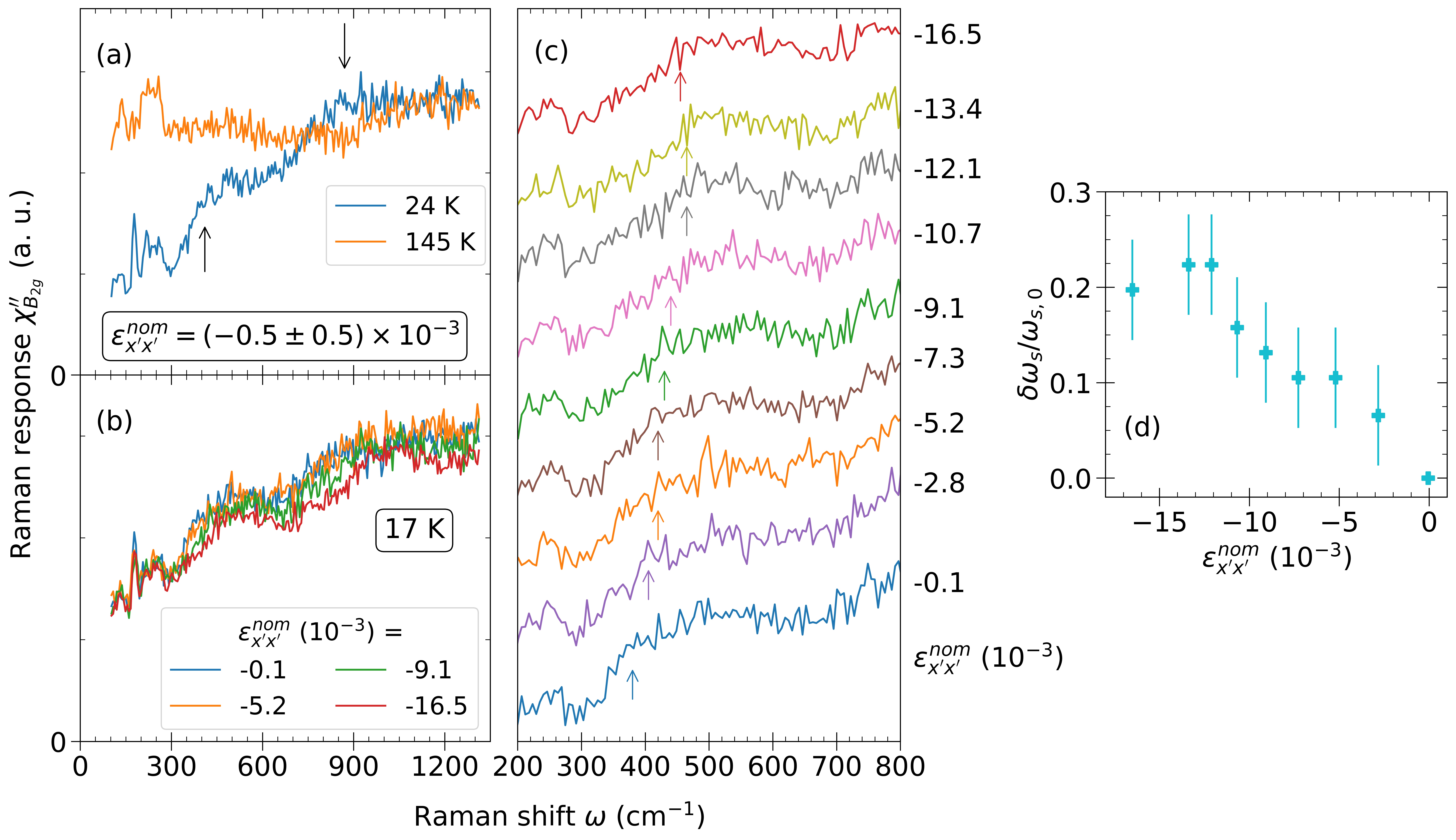}
    \caption{(a) Effect of the opening of the SDW gap on the Raman response without applied strain. The black arrows point to the two salient traits assigned to magnetism: the step and the broad peak, respectively at 400 and 850~cm$^{-1}$. (b) and (c) Strain effect on the SDW features. Panel (c) is a zoom on the spectra displayed in panel (b), with a vertical shift of the spectra to enhance clarity. The arrows point to the slope discontinuity at the step. (d) Relative change of the pinpointed $\omega_s$ with strain.}
    \label{figSDW}
\end{figure*}

In this section we present the effect of anisotropic strain on the SDW gap in BaFe$_2$As$_2$. The results are presented on Figure~\ref{figSDW}.

The magnetic order is an antiferromagnetic spin-density wave (SDW) order setting up in BaFe$_2$As$_2$ below $T_N = 138$~K. The opening of the SDW gap produces characteristic features on the zero strain Raman spectra~\cite{chauviere_raman_2011}, which we see on our low strain spectra (Fig.~\ref{figSDW} - (a)). The loss of spectral weight below 700~cm$^{-1}$ is a combination of both nematic and magnetic order effects. The specific traits of the opening of the SDW gap are the step-like feature around 350--450~cm$^{-1}$ and the broad peak developping around 700--1000~cm$^{-1}$.

Fig.~\ref{figSDW} - (b) and (c) report the effect of anisotropic strain (the same as in the main text) at 17~K. We observe a frequency shift of both the step and the peak, both hardening when strain increases. Note that at this low temperature, the quasi-elastic peak from the nematic fluctuations is hardly observed and the nematic susceptibility is significantly reduced, which explains why, as at 118~K, we observe no modification of the low energy response with strain other than that of the SDW feature.

To try and quantify this effect, we define $\omega_{s}$ as the step energy, that we pinpoint at the slope discontinuity marking the signal loss at the step. At low strain, this discontinuity is located at around 380~cm$^{-1}$, whereas at maximal compression we locate it at around 450~cm$^{-1}$, which corresponds to a 20~\% hardening at $\epsilon_{x^{\prime}x^{\prime}}^{nom} = -16.5 \times 10^{-3}$. The dependence of $\omega_s$ (renormalized to its zero-strain value) with strain is depicted in Fig.~\ref{figSDW} - (d). The 10~\% change of $\omega_s$ when $\epsilon_{x^{\prime}x^{\prime}}^{nom}$ is set to $-5\times 10^{-3}$~is consistent with the change in $T_N$ measured through nuclear magnetic resonance studies~\cite{kissikov_uniaxial_2018}.

This strain effect on the SDW gap shows that anisotropic strain tends to strengthen the magnetic order, consistently with previous nuclear magnetic resonance results ~\cite{kissikov_uniaxial_2018}. We highlight however that the strains depicted on Fig.~\ref{figSDW} correspond to rather high negative strains, up to $1.6 \times 10^{-2}$ (in absolute value). In particular, there is no visible effect for strains lower (in absolute value) than $3 \times 10^{-3}$, whereas the effect on the nematic fluctuations in this low strain regime is strong. This further confirms that strain only acts indirectly on magnetic order.

\subsection*{\label{sec:offset effect} Effect of an offset on the symmetrical behavior of the nematic scaling}
\begin{figure*}
\centering
  \subfloat{
	\begin{minipage}[c][1\width]{
	   1\textwidth}
	   \centering
	   \includegraphics[width=0.49\textwidth]{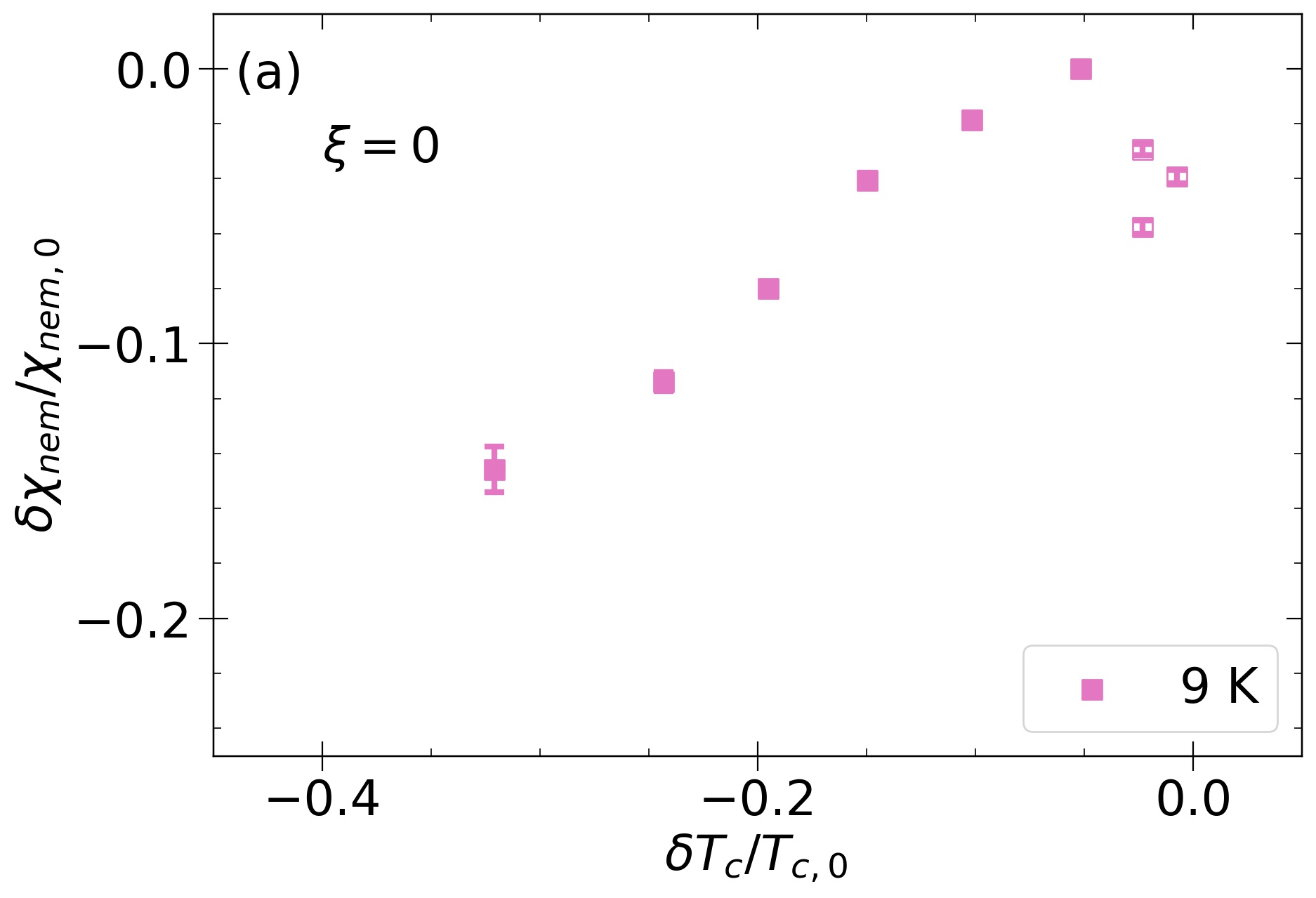}
	   \includegraphics[width=0.49\textwidth]{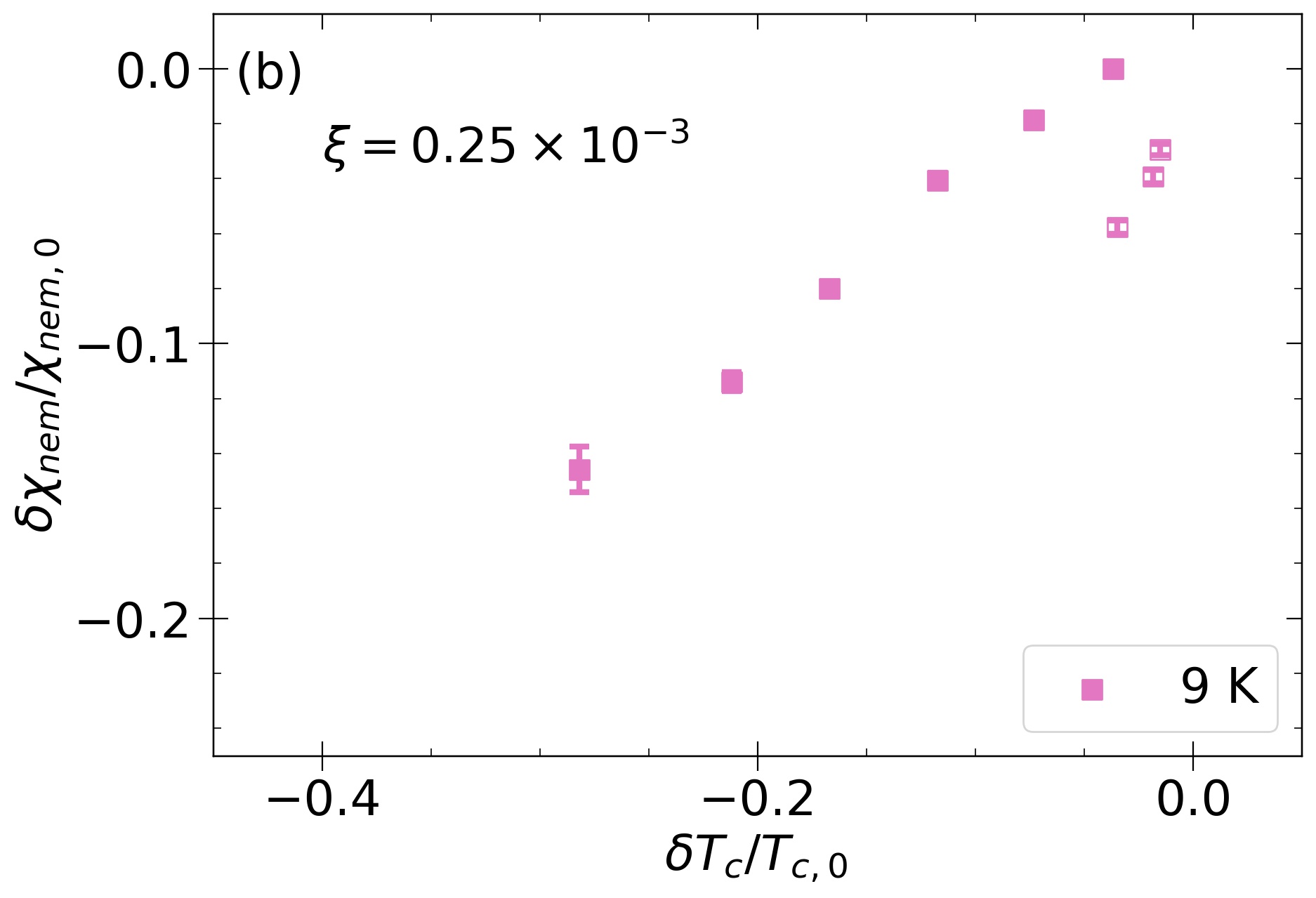}
	   \includegraphics[width=0.49\textwidth]{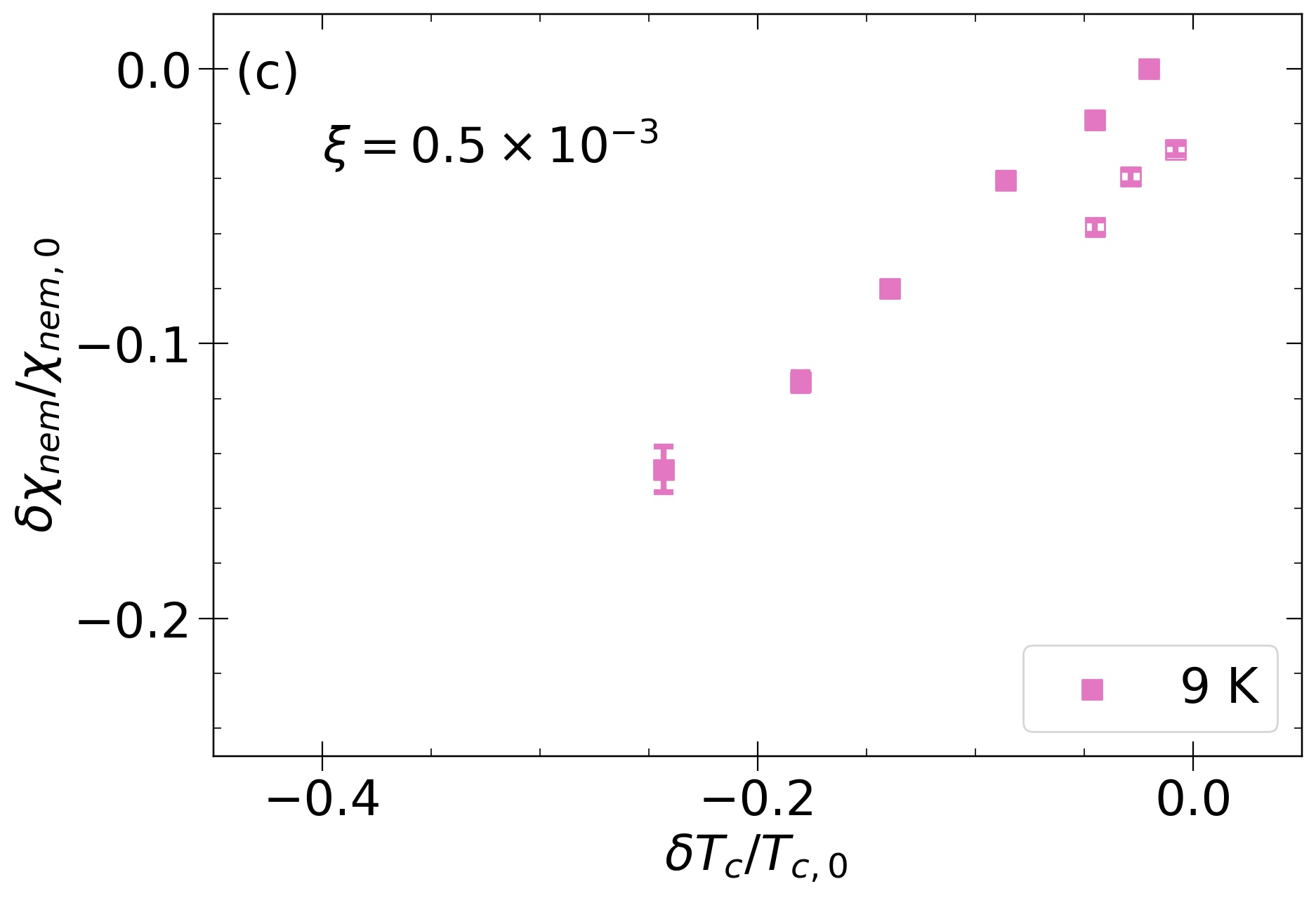}
	   \includegraphics[width=0.49\textwidth]{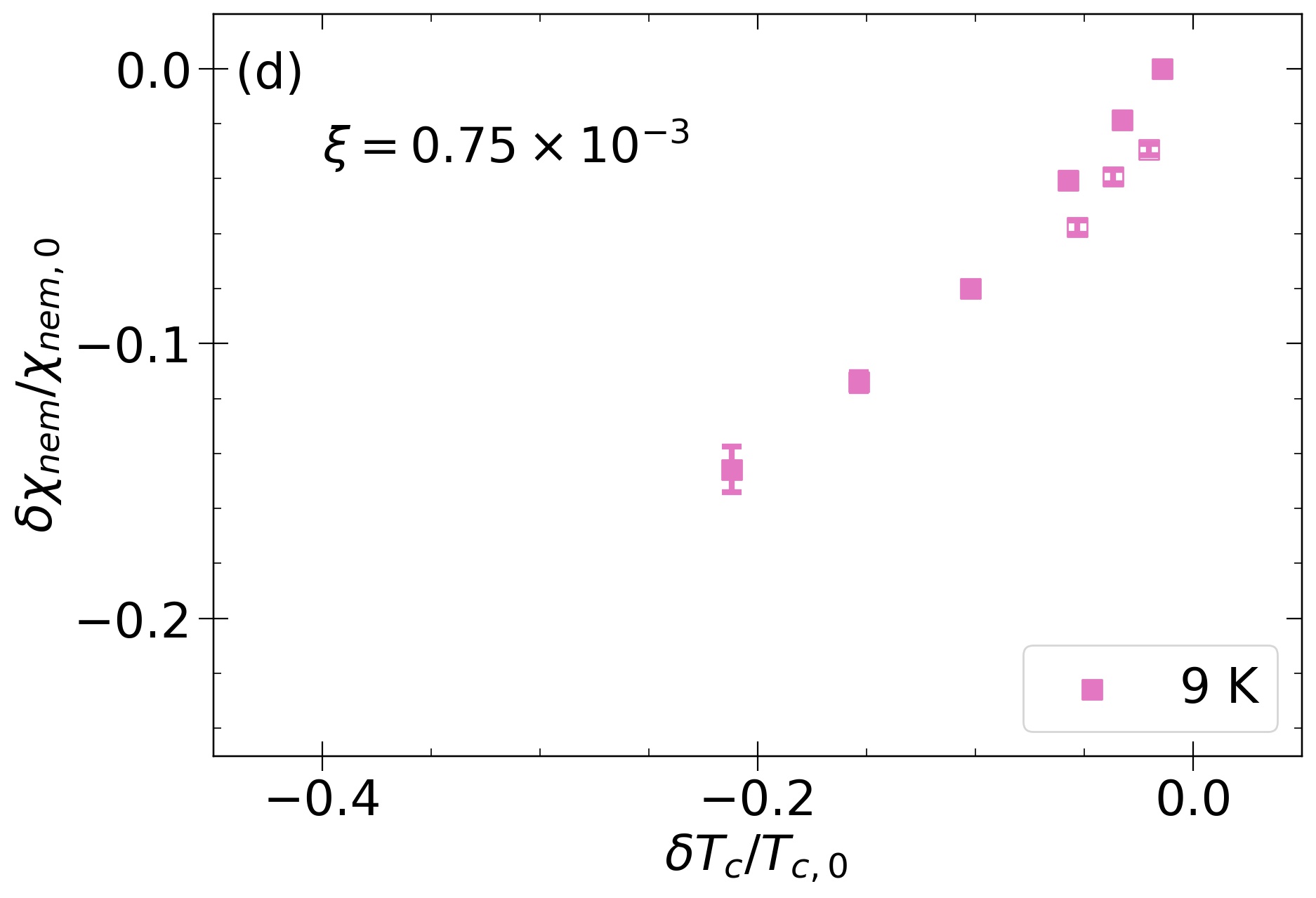}
	\end{minipage}}
    \caption{Effect of the choice of the offset $\xi$ of the nominal zero strain point on the asymmetry in the scaling btween positive and negative data points. $\xi$ is chosen as 0, 0.00025, 0.0005 and 0.00075 respectively for panels (a), (b), (c), (d). The empty symbols stand for positive strain, the filled ones for negative strain.}
    \label{fig:offset}
\end{figure*}

In Figure 3 of the main text, we show the scaling between the suppression of the nematic susceptibility and the decrease of $T_c$ under anisotropic strain. This scaling displays an asymmetry, with a clear offset between the negative strain data points and the fewer positive strain data points, both at 9 and 26~K. Even though an asymmetry could be expected through various effects (e.g. the effect of the $A_{1g}$ strain component), we think that the observed asymmetry is mainly due to the experimental uncertainty in the zero strain position. Indeed, it can slightly differ from the nominal zero strain position of the strain cell because of plastic deformation of the glue and thermal cycling. Considering the maximum of $\chi_{nem}$ as a good property of the zero strain point, we had to add an offset $\xi$ on the strain of + 0.5 $\times$ 10$^{-3}$ in our analysis displayed in Figures 2 and 3 of the main text. 
In this section, we show that the offset observed in Figure 3 between positive and negative strain depends greatly on the choice of the offset. To illustrate this point, in Fig.~\ref{fig:offset}, we plot the same quantities as in Figure 3 of the main text for the 9~K data, but we change the offset $\xi$ from 0 to + 0.75 $\times$ 10$^{-3}$ from panel (a) to panel (d): within this range of offset, the maximum of $\chi_{nem}$ stays very close to nominal zero strain point. So we conclude that in our data, the strong asymmetry between positive and negative strain in the scaling between the suppression of $\chi_{nem}$ and the decrease of $T_c$ is likely due to the uncertainty in locating the zero-strain state.

\subsection*{\label{sec:lambda}Doping dependence of the nemato-elastic coupling}

In nematic materials displaying superconductivity like Co:Ba122, nematic fluctuations are a possible origin for the enhancement of superconducting critical temperature near the nematic QCP~\cite{lederer_enhancement_2015}. However, the coupling between the nematic and the lattice degrees of freedom can quench critical nematic fluctuations near the nematic QCP, thus limiting its impact on the increase of $T_c$~\cite{labat_pairing_2017}. The nemato-elastic coupling is measured through the coupling constant $\lambda$, which appears in the Landau-Ginzburg expansion of Eq.~\eqref{eq:freeEnergy}.

$\lambda$ has been experimentally assessed through several techniques. However, it is difficult to extract quantitative absolute values of $\lambda$, and moreover, theoretical works do not give precise values of $\lambda$ beyond which nematic fluctuations can be quenched. In other words, the quantitative link between $\lambda$ and $T_c$ is currently not theoretically established. A way to nevertheless address the issue of the quenching of the nematic fluctuations through the nemato-elastic coupling is to evaluate the doping dependence of $\lambda$: a strong dependence approaching the QCP is a hint of its role on the critical nematic fluctuations. Through the relation of $\lambda$ to the structural transition temperature $T_s$ and the bare nematic transition temperature $T_0$ measured through the elastic coefficients~\cite{bohmer_nematic_2014} or Raman spectroscopy~\cite{gallais_observation_2013,gallais_charge_2016}, it was obtained that in Co:Ba122 $\lambda$ hardly changes between $x=0$ and the QCP doping. However recently through elastocaloric measurements, Ikeda et al.~\cite{ikeda_elastocaloric_2020} obtained a drastic suppression of $\lambda$ as doping increases, with a decrease by a factor of about 5. Thus no clear conclusion can be drawn from previous experimental results.

Through our elasto-Raman scattering measurements, we can also address this issue by comparing the suppression of the nematic fluctuations at the two probed doping levels. Indeed, as it appears in Eq.~(3) of the main text, the relative variation of nematic susceptibility with strain goes as $\epsilon_{x^{\prime}x^{\prime}}^2$, with $\lambda^2$ appearing in the prefactor.

\begin{figure}
    \centering
    \includegraphics[width=0.49\textwidth]{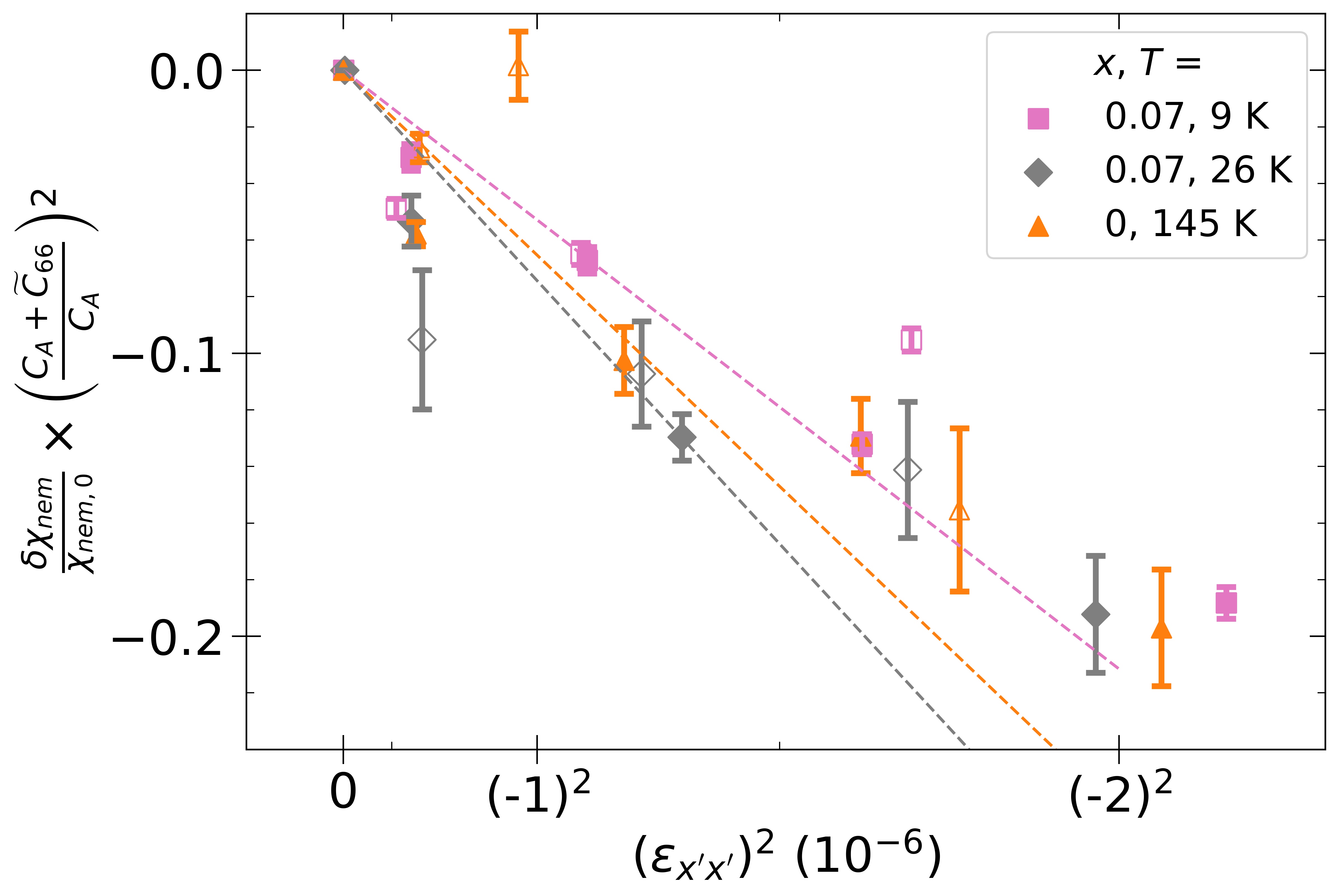}
   \caption{Comparison of the strain effect on $\chi_{nem}$ for the two doping levels. The empty symbols stand for the data under positive strain, the filled ones under negative strain. The dashed lines are guides to the eye following quadratic laws at low strain, with slopes of around $-6.5\times10^4$, $-5.4 \times10^4$ and $-7.5\times10^4$ respectively for $x=0$ at 145~K, $x=0.07$ at 9~K, and $x=0.07$ at 26~K.}
    \label{figLambda}
\end{figure}

In Figure~\ref{figLambda}, we compare the suppression of nematic fluctuations under strain in the two samples. To carefully compare the two samples, we renormalize the relative variation of susceptibility by the elastic coefficients factor $[C_A / (C_A + \tilde{C}_{66})]^2$ (note that $\tilde{C}_{66}$ is the experimentally measured shear modulus, which is renormalized with respect to the bare shear modulus $C_{66}$ through the nemato-elastic coupling). For $x=0$ at 145~K and $x=0.07$ at both 9~K and 26~K, we took $C_A$ respectively equal to 55.5 and 70~GPa, and $\tilde{C}_{66}$ respectively equal to 5 and 20~GPa, from data by Fujii et al.~\cite{fujii_diverse_2018}. Also, for the abscissa axis we consider $\epsilon_{x^{\prime}x^{\prime}}$ and not $\epsilon_{x^{\prime}x^{\prime}}^{nom}$ using the transmission ratio $\epsilon_{x^{\prime}x^{\prime}} = 0.82 \epsilon_{x^{\prime}x^{\prime}}^{nom}$ for the $x=0.07$ at 9K and 26K, and $\epsilon_{x^{\prime}x^{\prime}} =  \epsilon_{x^{\prime}x^{\prime}}^{nom}$ for $x$=0 at 145 K (near $T_s$). By adopting a quadratic scale for the abscissa axis, the theoretical slope is equal to $-12 b \lambda^2 \chi_{nem,0}^{3}$.

Despite a relative scatter in the data points, it is clear that the slopes do not appear to vary significantly with $x$. We note that $\chi_{nem,0}$ varies little between $x=0$ and $x=0.07$ with at most a 1.5 factor decrease, but as it appears to the power 3, even small changes can quantitatively impact the slope. Considering the extreme cases of no doping dependence for $\chi_{nem,0}$ or a 1.5 factor decrease, we obtain for $\frac{\sqrt{b}\lambda(x=0.07)}{ \sqrt{b}\lambda(x=0)}$ a maximum of 2.

Thus, unless the quartic coefficient $b$ increases significantly with doping which we believe is unlikely, our results tend to show that $\lambda$ is not strongly suppressed towards the nematic QCP in Co:Ba122.

\end{document}